\documentclass[10pt,final,doublecolumn]{IEEEtran}
\usepackage{balance}
\usepackage{amsthm}
\usepackage{amsmath}
\usepackage{latexsym}
\usepackage{graphicx}
\usepackage{bbding}
\usepackage{indentfirst}
\usepackage{algorithm,algorithmic}
\usepackage{setspace}
\usepackage{float}
\usepackage{epstopdf}
\usepackage{amssymb}
\usepackage{amsfonts}
\usepackage{enumerate}
\usepackage{multicol}
\usepackage{color}
\usepackage{diagbox}
\usepackage{bm}
\usepackage{amssymb}
\usepackage{stfloats}
\usepackage{epstopdf}
\usepackage{threeparttable}
\usepackage{epstopdf}
\usepackage{threeparttable}
\usepackage{makecell}
\usepackage{cite}
\usepackage{graphicx}
\usepackage{subcaption}

\IEEEoverridecommandlockouts
\allowdisplaybreaks[4]

\begin{document}
	\title{Polarforming Antenna Enhanced Sensing and Communication: Modeling and Optimization}

	\author{{Xiaodan Shao, \IEEEmembership{Member, IEEE}, Rui Zhang, \IEEEmembership{Fellow, IEEE}, Haibo Zhou, \IEEEmembership{Senior Member, IEEE}, Qijun Jiang, Conghao Zhou, \IEEEmembership{Member, IEEE}, Weihua Zhuang, \IEEEmembership{Fellow, IEEE}, Xuemin (Sherman) Shen, \IEEEmembership{Fellow, IEEE}}\vspace{-15pt}
		\thanks{X. Shao, C. Zhou, W. Zhuang, and X. Shen are with the Department of Electrical and Computer Engineering, University of Waterloo, Waterloo, ON N2L 3G1, Canada (E-mail: x6shao@uwaterloo.ca, c89zhou@uwaterloo.ca,
	wzhuang@uwaterloo.ca, sshen@uwaterloo.ca).}	
\thanks{H. Zhou is with the School of Electronic Science and Engineering, Nanjing University, Nanjing 210023,
	China (E-mail: haibozhou@nju.edu.cn).}
\thanks{Q. Jiang is with School of Science and Engineering,  The Chinese University of Hong Kong, Shenzhen, Guangdong 518172, China (E-mail: qijunjiang@link.cuhk.edu.cn).}
\thanks{R. Zhang is with School of Science and Engineering, Shenzhen Research Institute of Big Data, The Chinese University of Hong Kong, Shenzhen, Guangdong 518172, China. He is also with the Department of Electrical and Computer Engineering, National University of Singapore, Singapore 117583 (E-mail: elezhang@nus.edu.sg).}}
	\maketitle
	
	\IEEEpeerreviewmaketitle
	
	\begin{abstract}
In this paper, we propose a novel {\emph{polarforming antenna (PA)}} to achieve  cost-effective wireless sensing and communication. Specifically, the PA can enable polarforming to adaptively control the antenna's polarization electrically as well as tune its position/rotation mechanically, so as to effectively exploit polarization and spatial diversity to reconfigure wireless channels for improving sensing and communication performance. To analyze the performance gain of PA, we study an PA-enhanced integrated sensing and communication (ISAC) system that utilizes user location sensing to facilitate communication between an PA-equipped base station (BS) and PA-equipped users, by focusing on a new practical channel setup where the locations of users are nearly time-invariant but their orientations may change frequently (e.g., mobile phones rotated by spectators seated in a stadium while taking live photos). First, we model the PA channel in terms of transceiver antenna polarforming vectors and antenna positions/rotations.
We then propose a two-timescale ISAC protocol, where in the slow timescale, user localization is first performed, followed by the optimization of the BS antennas' positions and rotations based on the sensed user locations; subsequently, in the fast timescale, transceiver polarforming is adapted to cater to the instantaneous orientation of user devices in three-dimensional (3D) space, with the optimized BS antennas' positions and rotations.
We propose a new polarforming-based user localization method that uses a structured time-domain pattern of pilot-polarforming vectors to extract the common stable components in the PA channel across different polarizations based on the parallel factor (PARAFAC) tensor model.
Moreover, we maximize the achievable average sum-rate of users by jointly optimizing the fast-timescale transceiver polarforming, including phase shifts and amplitude variations, along with the slow-timescale antenna rotations and positions at the BS.
Simulation results validate the effectiveness of  polarforming-based localization algorithm and demonstrate the performance advantages of polarforming, antenna placement, and their joint design in comparison with various benchmarks without polarforming or antenna position/rotation adaptation.
\end{abstract}
	\vspace{-3pt}	
	\section{Introduction}
	\vspace{-3pt}	
Integrated sensing and communication (ISAC) has emerged as a key
technology for the future/next six-generation (6G) wireless networks \cite{isac1,isac2}. By sharing
hardware platforms and signal processing modules, ISAC
enables efficient resource utilization for both communication
and sensing, improving spectral efficiency and reducing
hardware costs. However, conventional ISAC systems employ fixed-position-and-polarization antenna (FPPA) in sensing and communication \cite{wangzhe,haiquan1,pol1,duala,qingisac,proc}. While programmable waveforms and/or adaptive transceiver beamforming with a large antenna array can improve sensing accuracy and channel capacity, wireless channels still lack flexible control via antenna polarization and position/rotation adaptations.
When a signal propagates through the wireless channel, it may suffer random depolarization due to
scattering and other environmental effects \cite{pol1}. As a result, polarization mismatch occurs when there is a misalignment between the polarization orientations of the transmitting and receiving antennas, leading
to signifcant signal attenuation as the misaligned component
of the electromagnetic wave is severely attenuated. Conventional FPPA systems fail to adapt to channel variations caused by depolarization.
On the other
hand, as the locations of mobile users or targets in wireless
networks change over time, the spatial degrees of freedom
(DoFs) of FPPAs remain limited and cannot efficiently adapt to the dynamic spatial distribution of users in the network \cite{wangzhe}.
Notably, as each antenna at transmitter/receiver rotates, its polarization state changes accordingly. Therefore, antenna polarization and its position/rotation are tightly coupled, both of which need to be considered in wireless communication/sensing system designs.

\begin{figure}[t]
	\centering
	\setlength{\abovecaptionskip}{0.cm}
	\includegraphics[width=3.6in]{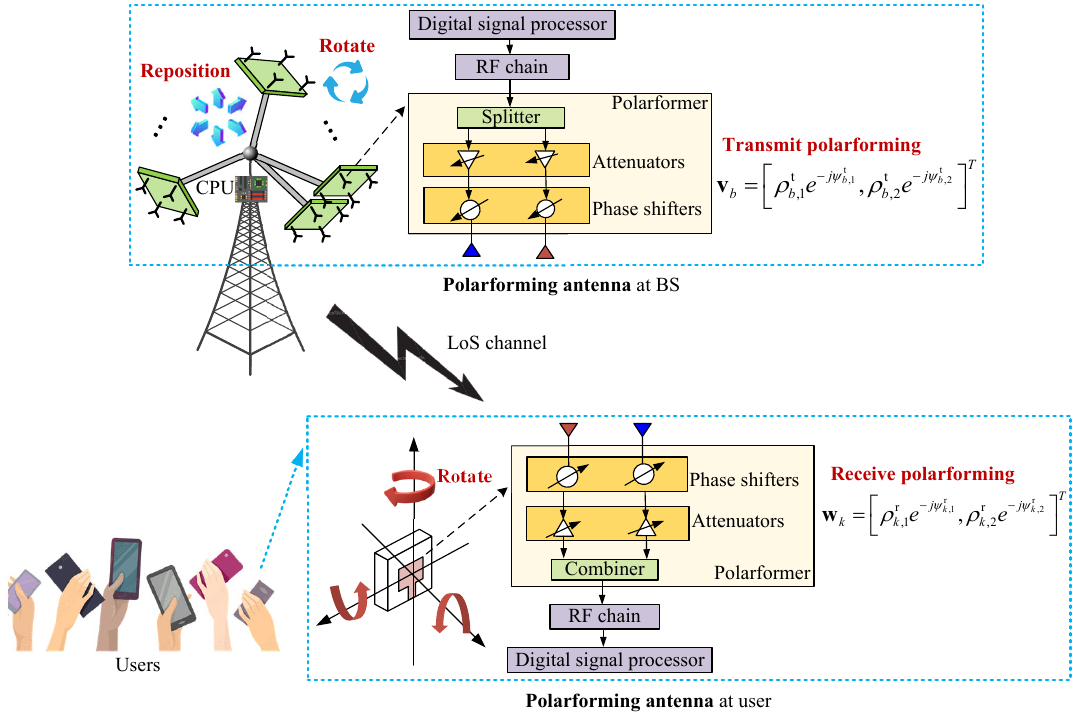}
	\caption{Illustration of the PA-enhanced sensing and communication system.}
	\label{practical_scenario}
	\vspace{-0.69cm}
\end{figure}

To enable the transceiver's full flexibility in antenna polarization and placement, we propose a novel polarforming antenna (PA), also called polarized six-dimensional movable antenna (6DMA) \cite{6dmatwc}, as a new approach to improve the performance of wireless sensing and communication without the need to add more antennas and bear their additional  cost and energy consumption. 
Specifically, an PA tunes the polarforming vector with adjustable signal amplitude and phase to control the polarization of each antenna at the transmitter and/or receiver, thus allowing dynamic adjustment of antenna polarization in response to real-time channel conditions and depolarization effects by exploiting polarization diversity. This is achieved through the electronically tunable polarformer, which contains phase shifters and attenuators that collectively adjust the phase and amplitude of the transmitted/received signal (see Fig. \ref{practical_scenario}). 
Furthermore, each PA can be independently adjusted in terms of position and/or rotation to modify its polarization angle as well as adapt to
the channel spatial distribution.
This can be realized through mechanical control, which utilizes external mechanical structures with the aid of actuators, such as electric motors and precision gears, to achieve physical movement (see Fig. \ref{practical_scenario}).  

It is worth noting that the PA system proposed in this paper differs significantly from our recently proposed 6DMA \cite{6dmatwc,6dmaDis,6dmaMag} and traditional polarization reconfigurable antenna \cite{poa}.
Firstly, a mechanically controlled 6DMA, consisting of multiple three-dimensional (3D) rotatable and positionable antennas/subarrays, either lacks the antenna polarization control or is designed under a fixed polarization condition \cite{shao2025tutorial, jiang2025statistical, pi20246d, liu2024uav,10918750}. Secondly, although traditional polarization-reconfigurable antennas can adjust antenna polarization, they tune only the amplitude of the signal, which does not fully exploit polarization diversity. Moreover, polarization reconfigurable antennas with fixed antenna position/rotation cannot leverage antenna movement to achieve additional spatial DoFs. In contrast, employing the polarformer presented in \cite{ppp}, the proposed PA aims at collectively adjusting both the amplitude and phase of the signals to fully explore the benefits of polarization diversity, as well as adjusting its antenna position/rotation to enhance spatial DoFs. A summary of the above comparison is given in Table I. 

Besides enhancing communication performance, the proposed PA is to help improve wireless sensing/localization performance. Various wireless localization systems exist, such as those based on WLAN or cellular networks, which can determine the location of a user or target object using dedicated wireless reference signals between reference nodes and target nodes \cite{isac1,shaos,nsr,yumeng}. A target object at an unknown location is often called agent node, while a reference node with a known position is referred to as anchor node. In some cases, the agent node determines its location by receiving reference signals from anchor nodes. However, there are other cases where the anchors receive reference signals from agents to localize the agents, and then inform them the estimated locations via a separate communication link \cite{liuan}. Here, we consider the latter case, where multiple PAs are deployed at the anchor/base station (BS) to receive reference signals from single-PA users for user localization.

\begin{figure}[t]
	\centering
	\setlength{\abovecaptionskip}{0.cm}
	\includegraphics[width=3.5in]{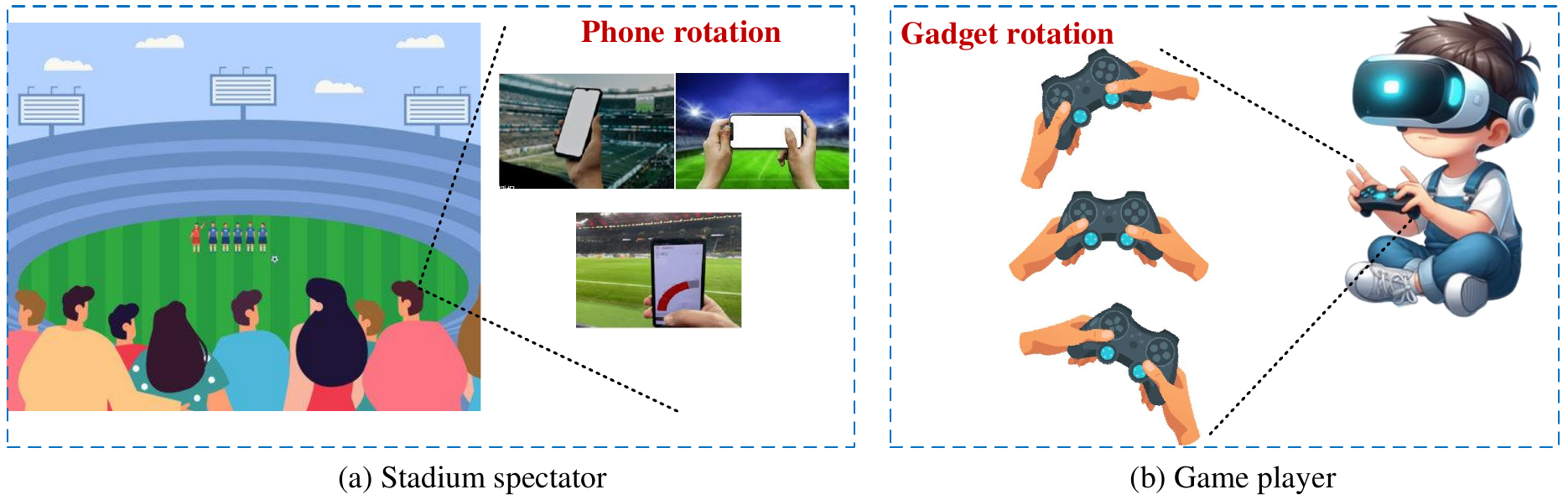}
	\caption{Examples of user rotation scenarios in PA-enhanced ISAC systems.}
	\label{userrotate}
\vspace{-0.69cm}
\end{figure}

Motivated by the above, we consider an PA-enhanced ISAC system that utilizes
user location sensing to facilitate communication between PA-equipped users and an PA-equipped BS. In our considered system, the transmit and receive polarforming vectors, along with the positions and rotations of the PA subarrays at the BS, need to be jointly designed to achieve optimal sensing and communication performance. However, the design of these parameters depends on the user's position and orientation. Thus, we consider a practical two-timescale channel setup, where the users' positions change slowly and can be assumed to be fixed within a sufficiently long period termed as location coherence time, while their orientations may vary frequently during this period (intentionally or non-intentionally). This channel setup is applicable to many scenarios, such as spectators seated in a stadium watching a football match, where their smartphones remain stationary in position but are frequently rotated to take photos in different directions (see Fig. \ref{userrotate} (a)). Another example is for users playing virtual reality (VR) games at fixed locations, where their VR gadgets rotate frequently to enhance the gaming experience (see Fig. \ref{userrotate} (b)). 
\begin{table*}[!t]
		\vspace{-0.5cm}
\footnotesize
	\caption{Comparison of PA with Other Related Technologies.}
	\label{Table1}
	\centering
	\begin{tabular}{|c|c|c|c|c|c|}
		\hline
		\makecell{Technology} & \makecell{Polarization adaptability} & \makecell{Adaptability to channel \\ spatial distribution} & \makecell{Control methods} & \makecell{Performance \\ Gain} & \makecell{Hardware \\ Cost} \\
		\hline
		\makecell{Polarforming antenna (PA)} & \makecell{High \\ (Both phase and \\ amplitude-adjustable polarforming)} & \makecell{High \\ (rotation and position)} & \makecell{Electrical \\ and mechanical}  & \makecell{Very high} & \makecell{Medium} \\
		\hline
		6DMA \cite{6dmatwc,6dmaDis} & No & \makecell{High \\ (rotation and position)}  & Mechanical & High & Medium \\
		\hline
		\makecell{Polarization reconfigurable \\ antenna \cite{poa}} & \makecell{Low \\ (only amplitude-adjustable polarization)} & No & Electrical & Moderate  & Low \\
		\hline
	\end{tabular}
		\vspace{-0.3cm}
\end{table*}   

In the proposed PA-enhanced ISAC system, the positions and rotations of antennas are mechanically controlled, which may have limited movement speeds, rendering them more suitable for adapting to slow-timescale channel variations, e.g., user locations. In contrast, transmit and receive polarforming of PA is electronically controlled, thus allowing their rapid adjustments to adapt to fast-timescale channel variations caused by, e.g., users' frequent rotations. To design such a system, the acquisition of channel state information
(CSI) is essential. However, since the BS can only measure user-BS channels at the positions and rotations where BS's subarrays are physically deployed, it is difficult to obtain the instantaneous CSI across the entire antenna movement region at the BS and for arbitrary users' positions and rotations. Fortunately, in our considered scenarios (see Fig. \ref{userrotate}), the user's locations remain unchanged within each location coherence time, which can be acquired and used for the optimization of BS antenna positions/rotations. Then, with the optimized antenna positions/rotations at the BS, the instantaneous CSI of users due to their rotations can be measured for designing the fast-timescale transmit and receive polarforming.     
The main contributions of this paper are summarized as follows. 
\begin{itemize}
	\item
First, we propose the new PA to electrically adjust antenna polarization through polarforming vectors to align the transmitter and receiver polarizations, as well as mechanically adjust antenna position/rotation to adapt to user spatial channel distribution. Moreover, we model the PA channel and decouple the channel into stable unpolarformed components and dynamic polarformed components, which provides essential insights into our considered channel setup and  facilitates the determination of parameters in the PA-enhanced system.    
    
 \item  Next, we study an PA-enhanced ISAC system that utilizes user location sensing to facilitate communications between PA-equipped users and an PA-equipped BS. For the new practical channel setup where user locations are nearly static but orientations may vary frequently, we propose a two-timescale ISAC protocol. In the slow timescale, user localization is first performed, based on which the BS antennas' positions and rotations are optimized, while subsequently in the fast timescale, transceiver polarforming is adapted to cater to the instantaneous user/device rotations in 3D space.

	\item 
Moreover, we design a new polarforming-based user localization method that leverages controllable polarforming vectors to create a time-domain pattern of pilot-polarforming vectors. This enables the extraction of stable line-of-sight (LoS) components in the PA channel across different polarizations using the parallel factor (PARAFAC) tensor. Thereby, user locations can be accurately determined from the estimated LoS components. Simulation results validate the effectiveness of the proposed localization method by exploiting the new DoF of polarforming.
		
	\item
Finally, we formulate an optimization problem to maximize the weighted sum-rate of users by jointly designing transceiver polarforming and BS antenna positions/rotations based on sensed user locations. To solve this non-convex problem, we propose an efficient algorithm that accounts for PA movement constraints, discrete polarforming amplitude and phase shifts, and enables efficient variable updates via closed-form solutions or simple iterations. We show by simulation that polarforming optimization improves the achievable rate over fixed polarization, with more gains achieved when being jointly optimized with antenna position/rotation.
\end{itemize}

The rest of this paper is organized as follows. Section II presents the PA architecture and the corresponding channel model. Section III presents a two-timescale protocol design for the PA-enhanced ISAC system. Section IV presents the polarforming-based localization algorithm. Section V addresses antenna polarforming and position/rotation optimization. Section VI presents simulation results for performance evaluation. Finally, Section VII concludes this study.

	\emph{Notations}: Symbols \((\cdot)^*\), \(\dagger\), \((\cdot)^H\), and \((\cdot)^T\) denote the operations of conjugate, inverse, conjugate transpose, and transpose, respectively, \(\mathbb{E}[\cdot]\) represents the expectation over a random variable, \(\mathbf{I}_N\) denotes the \(N \times N\) identity matrix,  \(\lceil\cdot\rceil\) denotes the ceiling operator, \(\mathcal{O}(\cdot)\) denotes big-O notation, \(\cup\) represents set union, \(\mathbf{a} \cdot \mathbf{b}\) denotes the dot product of vectors \(\mathbf{a}\) and \(\mathbf{b}\), \(\otimes\) and \(\circ\) denote the Kronecker and Khatri-Rao  matrix products, respectively, \(\|\cdot\|\), $\|\cdot\|_F$, and \(\|\cdot\|_\infty\) denote the Euclidean norm, Frobenius norm, and infinity norm of a complex vector, respectively, \(\mathcal{CN}(0, \sigma^2)\) defines a circularly symmetric complex Gaussian (CSCG) distribution with mean 0 and variance \(\sigma^2\), \(|\mathcal{X}|_{\mathrm{c}}\) represents the cardinality of set \(\mathcal{X}\), \(\mathrm{diag}(\mathbf{a})\) forms a diagonal matrix with vector \(\mathbf{a}\) as its diagonal elements, and \([\mathbf{a}]_j\) refers to the \(j\)-th element of \(\mathbf{a}\).
\vspace{-3pt}	
	\section{System Model} 
	\vspace{-3pt}	
	We consider an PA-enhanced ISAC system as shown in Fig. \ref{practical_scenario}, where multiple PAs are deployed at the BS to receive reference signals from single-PA users for user localization as well as to communicate with these users over a given frequency band. Without loss of generality, we consider the downlink communications from the BS to users.
	In this section, we first describe the new PA architecture and then present its corresponding channel model, which lays the foundation for the subsequent sensing/localization and communication designs.	
\vspace{-3pt}
\subsection{PA Architecture}
\vspace{-3pt}
As shown in Fig. \ref{practical_scenario}, each PA can independently induce a certain phase shift (via the attached phase shifters in the polarformer) and amplitude change (via the attached attenuators in the polarformer) on the transmit/receive signals, thereby adjusting the antenna's polarization. The phase shifters and attenuators in the polarformer enable precise polarization control by independently adjusting the phase and amplitude of signals. In this way, PA can enable polarforming to exploit polarization diversity and adaptively control the antenna polarization for aligning polarization between transmitting and receiving antennas. We assume that each transmit/receive antenna consists of two orthogonally oriented linearly polarized elements, with one element for vertical polarization (\(\mathcal{V}\)-element) and the other for horizontal polarization (\(\mathcal{H}\)-element).

Specifically, each user \(k\) is equipped with a single PA, and the receive polarforming vector is denoted by
	\begin{align}
		\mathbf{w}_k = \begin{bmatrix} 
			\rho_{k,1}^{\mathrm{r}}e^{-j \psi_{k,1}^{\mathrm{r}}}, \rho_{k,2}^{\mathrm{r}}e^{-j \psi_{k,2}^{\mathrm{r}}} \end{bmatrix}^T,
	\end{align}
		where \( \rho_{k,1}^{\mathrm{r}} \in[0,1]\) and \( \rho_{k,2}^{\mathrm{r}}\in[0,1] \) represent the amplitude coefficients for the \(\mathcal{V}\)- and \(\mathcal{H}\)-elements, respectively, of the corresponding antenna at the user. In addition, \( \psi_{k,1}^{\mathrm{r}} \in[0,2\pi)\) and \( \psi_{k,2}^{\mathrm{r}} \in[0,2\pi)\) represent the phase shifts for the user's \(\mathcal{V}\)- and \(\mathcal{H}\)-elements, respectively.
		
The BS has $B$ PA subarrays, denoted by $\mathcal{B} = \{1, 2, \ldots, B\}$. Each PA subarray  consists of 
$N$ $(\geq 1)$ PAs, denoted by $\mathcal{N} = \{1, 2, \ldots, N\}$. The total number of transmit antennas at the BS is thus $BN$. Each PA subarray is a uniform planar array (UPA) with a given size. All antennas within the same BS subarray share identical polarization characteristics determined by the propagation environment. Thus, their polarization can be controlled by the same polarforming vector. For the \(b\)-th PA subarray at the BS, the transmit polarforming vector is given by 
	\begin{align}
		\mathbf{v}_b = \frac{1}{\sqrt{2}}\begin{bmatrix} 
			\rho_{b,1}^{\mathrm{t}}e^{-j \psi_{b,1}^{\mathrm{t}}}, \rho_{b,2}^{\mathrm{t}}e^{-j \psi_{b,2}^{\mathrm{t}}} \end{bmatrix}^T,
	\end{align}
	where \( \rho_{b,1}^{\mathrm{t}} \in[0,1]\) and \( \rho_{b,2}^{\mathrm{t}}\in[0,1] \) represent the amplitude coefficients for the \(\mathcal{V}\)- and \(\mathcal{H}\)-elements, respectively, of each corresponding antenna at the $b$-th transmit PA subarray. Similarly, \( \psi_{b,1}^{\mathrm{t}}\in[0,2\pi)\) and \( \psi_{b,2}^{\mathrm{t}} \in[0,2\pi)\) represent the phase shifts for the \(\mathcal{V}\)- and \(\mathcal{H}\)-elements, respectively, for the $b$-th transmit PA subarray antennas. The phase and amplitude of the polarforming vector can be controlled either continuously or discretely. We consider the discrete control of amplitude and phase shifts for the ease of implementation. Let $Q_{\rho}$ and $Q_{\vartheta}$ denote the number of bits for polarforming amplitude and phase-shift control per antenna element, respectively. We thus have
		\begin{figure}[t!]
	\vspace{-0.5cm}
		\centering
		\setlength{\abovecaptionskip}{0.cm}
		\includegraphics[width=2.5in]{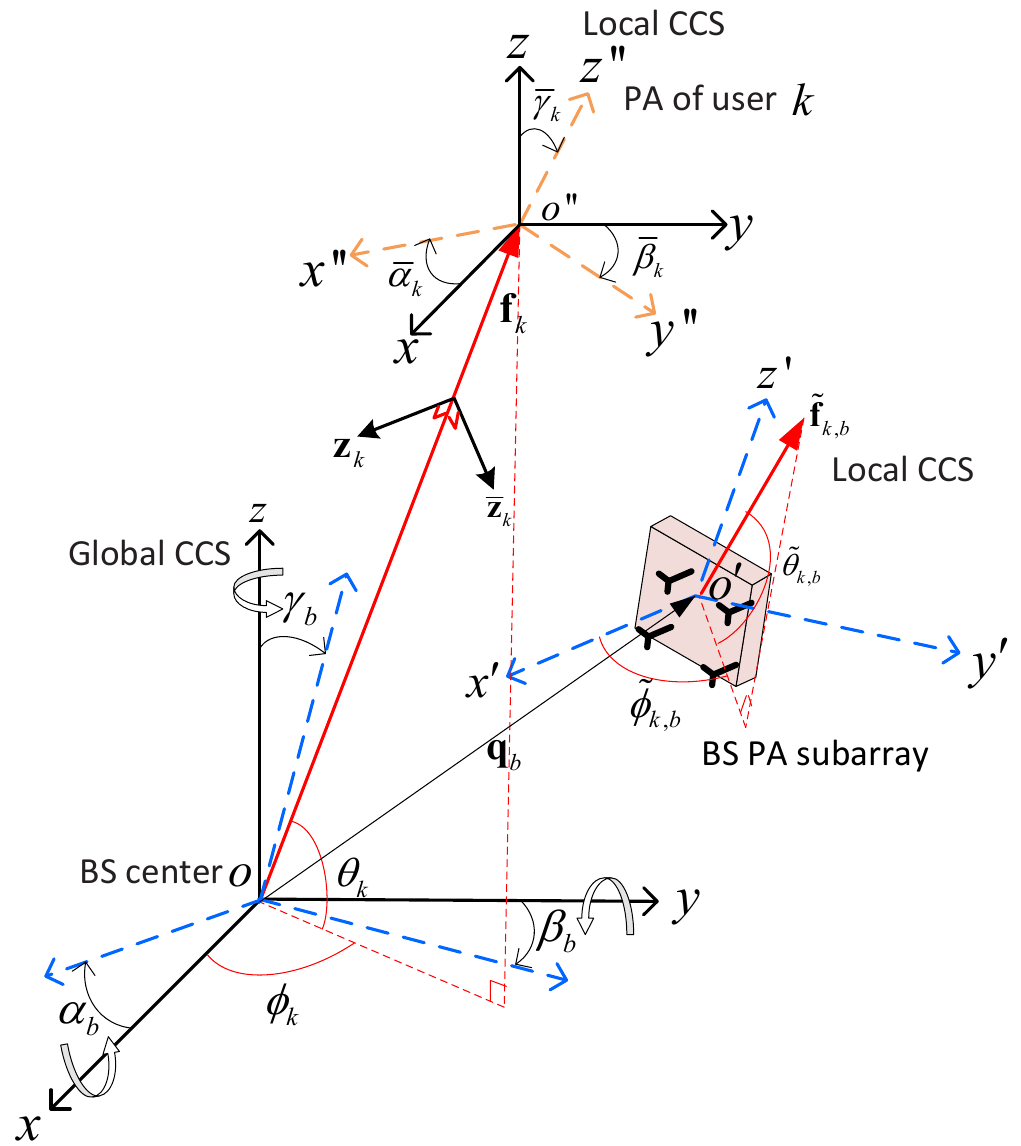}
		\caption{Geometry illustration of the $b$-th PA subarray at the BS and the PA at user $k$.}
		\label{system}
\vspace{-0.69cm}
	\end{figure}
\begin{align}  
	\!\!\!\![\mathbf{w}_k]_i \text{ and } [\mathbf{v}_b]_i \in \mathcal{F} \!\triangleq\! \{ \chi \mid  \chi = \rho e^{j \vartheta },  \vartheta \in \mathcal{S},  \rho \in \mathcal{A}\},   
\end{align}  
for $i=\{1,2\}$, where \(\mathcal{F}\) denotes the set of all possible values for \( \chi\), consisting of both amplitude \(\rho\) and phase \(\vartheta\),
$\mathcal{S} \triangleq \{0, \frac{2\pi}{D}, \ldots, \frac{2\pi(D-1)}{D}\}$ with $D = 2^{Q_{\vartheta}}$. Here, the discrete phase values are assumed to be equally spaced in the interval $[0, 2\pi)$, \(\mathcal{A} \triangleq \{{\rho}_1, \ldots, {\rho}_{2^{Q_{
			\rho}}}\}\) denotes the controllable amplitude set with \(|\mathcal{A}| = 2^{Q_{
		\rho}}\), and \(\rho_i, i = 1, \ldots, 2^{Q_\rho}\), are equally spaced in the interval \([0, ])\). Note that, when \(Q_{\rho} = 0\), \(\mathcal{A}\) reduces to the case without attenuation, i.e., \(\mathcal{A} = \{1\}\).

Each PA can rotate and/or reposition within a given space. We assume that each user PA can rotate while maintaining a fixed position. This is achieved with actuators or parasitic elements such as a rotating motor, or by manual adjustment, to enable physical movement. In contrast, all antennas on each BS PA subarray can reposition and rotate together within a given convex 3D space \(\mathcal{C}\) at the BS. This is mechanically controlled by connecting the subarrays to the central processing unit (CPU) of the BS via extendable and rotatable rods embedded with flexible wires, allowing the CPU to precisely control their 3D positions and rotations \cite{6dmatwc}.

To facilitate the user/BS antenna movement description, we establish three Cartesian coordinate systems (CCSs). As shown in Fig. \ref{system}, the global CCS is denoted by \(o\text{-}xyz\), with the BS's CPU (i.e., BS center) centered at  origin \(o\). Each PA subarray’s local CCS is denoted by \(o'\text{-}x'y'z'\), with the subarray center as  origin \(o'\). Each user's local CCS is denoted by \(o''\text{-}x''y''z''\), with origin \(o''\) taken as the center of the user's antenna.  
On the BS side, the position and rotation of the $b$-th PA subarray, $b\in\mathcal{B}$, can be characterized by the following position vector and rotation vector
	\begin{align}\label{bb}
		\mathbf{q}_b=[x_b,y_b,z_b]^T\in\mathbb{R}^{3 \times 1},~
		\mathbf{u}_b=[\alpha_b,\beta_b,\gamma_b]^T\in \mathbb{R}^{3 \times 1},
	\end{align}
 where $x_b$, $y_b$ and $z_b$ represent the coordinates of the $b$-th PA's center in the global CCS; $\alpha_b$,  $\beta_b$ and $\gamma_b$ all in \([0, 2\pi)\)  denote the rotation angles of the $b$-th PA subarray with respect to (w.r.t.) the $x$-axis, $y$-axis, and $z$-axis in the global CCS, respectively.
Given $\mathbf{u}_b$, the corresponding rotation matrix can be written as
	\begin{align}\label{R}
		&\!\!\!\mathbf{R}(\mathbf{u}_b)
		&\!=\!\begin{bmatrix}
			c_{\beta_b}c_{\gamma_b} & c_{\beta_b}s_{\gamma_b} & -s_{\beta_b} \\
			s_{\beta_b}s_{\alpha_b}c_{\gamma_b}-c_{\alpha_b}s_{\gamma_b} & s_{\beta_b}s_{\alpha_b}s_{\gamma_b}+c_{\alpha_b}c_{\gamma_b} & c_{\beta_b}s_{\alpha_b} \\
			c_{\alpha_b}s_{\beta_b}c_{\gamma_b}+s_{\alpha_b}s_{\gamma_b} & c_{\alpha_b}s_{\beta_b}s_{\gamma_b}-s_{\alpha_b}c_{\gamma_b} &c_{\alpha_b}c_{\beta_b} \\
		\end{bmatrix},\!\!
	\end{align}
 where $c_{x}=\cos(x)$ and $s_{x}=\sin(x)$ \cite{6dmatwc}. Let $\bar{\mathbf{r}}_{n}$ denote the position of the $n$-th antenna of the PA subarray in its local CCS. Then, the position of the $n$-th antenna of the $b$-th PA subarray in the global CCS can be expressed as
	\begin{align}\label{nwq}
		\mathbf{r}_{b,n}(\mathbf{q}_b,\mathbf{u}_b)=\mathbf{q}_b+\mathbf{R}
		(\mathbf{u}_b)\bar{\mathbf{r}}_{n},~n\in\mathcal{N},~b \in\mathcal{B}.
	\end{align}

Next, on the user side, we let 
\begin{align}\label{URR}
	\mathbf{u}_{k}^{\mathrm{r}}=[\bar{\alpha}_k,\bar{\beta}_k,\bar{\gamma}_k]^T\in \mathbb{R}^{3 \times 1}
\end{align}
denote the rotation angle vector of the user's local CCS \(o''\text{-}x''y''z''\) relative to \(o\text{-}xyz\), where \(\bar{\alpha}_k\), \(\bar{\beta}_k\), and \(\bar{\gamma}_k\) all in \([0, 2\pi)\) denote the rotation angles of the $k$-th user w.r.t. the $x$-axis, $y$-axis and $z$-axis in the global CCS, respectively. Similar to \eqref{R}, given \(\mathbf{u}_{k}^{\mathrm{r}}\), the corresponding rotation matrix can be denoted by \(\mathbf{R}(\mathbf{u}_{k}^{\mathrm{r}})\), which is omitted for brevity.
\vspace{-3pt}	
\subsection{PA Channel Model}
\vspace{-3pt}
For the purpose of exposition, we assume that the channel between the BS and each user is a far-field LoS channel. In addition, the users' positions change slowly and can be considered to be approximately constant within each location coherence time, while their rotations (orientations) may vary arbitrarily over time during this period.
Each user is equipped with an omnidirectional antenna. Let $\phi_k\in[-\pi,\pi]$ and $\theta_k\in[-\pi/2,\pi/2]$ denote the azimuth and elevation angles, respectively, of the signal from user $k$ arriving at the BS w.r.t. its center.
	The pointing vector corresponding to direction $(\theta_k, \phi_k)$ is thus given by
	\begin{align}\label{KM}
		\mathbf{f}_k=[\cos(\theta_k)\cos(\phi_k), \cos(\theta_k)\sin(\phi_k), \sin(\theta_k)]^T.
	\end{align}
	By combining \eqref{nwq} and \eqref{KM}, 
	the steering vector of the $b$-th PA subarray is  given by
	\begin{align}\label{gen}
		\mathbf{a}_{k,b}(\mathbf{q}_b,\mathbf{u}_b)\!=\! \left[e^{-j\frac{2\pi}{\lambda}
			\mathbf{f}_k^T\mathbf{r}_{b,1}(\mathbf{q}_b,\mathbf{u}_b)},
		\!\cdots,\! e^{-j\frac{2\pi}{\lambda}\mathbf{f}_k^T
			\mathbf{r}_{b,N}(\mathbf{q}_b,\mathbf{u}_b)}\right]^T,
	\end{align}
	where $\lambda$ denotes the carrier wavelength.
 Next, to determine effective antenna gain $g_{k,b}(\mathbf{u}_{b})$, we project \( \mathbf{f}_k \) onto the local CCS of the \( b \)-th PA subarray, denoted by
	\begin{align}
				\tilde{\mathbf{f}}_{k,b}=\mathbf{R}(\mathbf{u}_{b})^{-1}\mathbf{f}_k.
	\end{align}
	Then, we represent $	\tilde{\mathbf{f}}_{k,b}$ in the spherical coordinate system as
	$	\tilde{\mathbf{f}}_{k,b}=[\cos(\tilde{\theta}_{k,b})\cos(\tilde{\phi}_{k,b}), \cos(\tilde{\theta}_{k,b})\sin(\tilde{\phi}_{k,b}), \sin(\tilde{\theta}_{k,b})]^T$, where $\tilde{\theta}_{k,b}$ and $\tilde{\phi}_{k,b}$ represent the corresponding directions of arrival (DoAs) in the local CCS (see Fig. \ref{system}).
	Finally, effective antenna gain $g_{k,b}(\mathbf{u}_{b})$ of the \(b\)-th PA subarray along direction \((\tilde{\theta}_{k,b}, \tilde{\phi}_{k,b})\) in the linear scale is given by
	\begin{align}\label{gm}
		g_{k,b}(\mathbf{u}_{b})=10^{\frac{A(\tilde{\theta}_{k,b}, 	\tilde{\phi}_{k,b})}{10}},
	\end{align}
	where \(A(\tilde{\theta}_{k,b}, \tilde{\phi}_{k,b})\) denotes the effective antenna gain in dBi determined by the antenna's radiation pattern \cite{6dmac}. Let
	\begin{align}\label{uik}
		\mathbf{h}_{k,b}^{\mathrm{LoS}}(\mathbf{q}_b,\mathbf{u}_b)=		\sqrt{\nu_k}e^{-j\frac{2\pi d_k}{\lambda}}\sqrt{g_{k,b}(\mathbf{u}_b)}
		\mathbf{a}_{k,b}(\mathbf{q}_b,\mathbf{u}_b)\in \mathbb{C}^{N\times 1},
	\end{align}
which denotes the {\emph{unpolarformed LoS channel}} between the $b$-th PA subarray at BS and the $k$-th user. The term ``unpolarformed” here refers to the fact that the polarforming effect is not considered yet. In \eqref{uik}, $\nu_k=\epsilon_0d_k^{-2
	}$ is the free-space path loss, where $\epsilon_0$ represents the channel power at the reference distance $d_0=1$ meter (m) \cite{airs}, and $d_k>d_0$ denotes the distance between the $k$-th user's location and the BS center. 

In the local CCS, the vertical \(\mathcal{V}\)-element of the antenna is aligned along the positive \(y'\)- or \(y''\)-axis, while the horizontal \(\mathcal{H}\)-element is oriented along the positive \(x'\)- or \(x''\)-axis, with their unit vectors respectively given by
\begin{align}
\mathbf{e}_\mathrm{v} = [0,1,0]^T, \quad \mathbf{e}_\mathrm{h} = [1,0,0]^T.
\end{align}
Moreover, the polarization state of an electromagnetic (EM) wave can be described by any two orthogonal electric field components on the wavefront, represented by the following orthogonal unit vectors in the global CCS \cite{heap}
\begin{align}
\mathbf{z}_k = [s_{\theta_k} s_{\phi_k}, -c_{\theta_k}, s_{\theta_k} c_{\phi_k}]^T,~~~
\bar{\mathbf{z}}_k = [c_{\phi_k}, 0, -s_{\phi_k}]^T,
\end{align}
which are perpendicular to the signal's direction of propagation (see Fig. \ref{system}) and characterize the polarization state of the EM wave propagating along the LoS path.

The transmit field components of the LoS path are generated by projecting the transmit antenna's time-varying electric fields onto the LoS signal direction. The corresponding transformation is  given by
	\begin{align}
	\!\!\!\!\!	\mathbf{P}_{k,b}(\mathbf{u}_b) \!= \!
		\begin{bmatrix}
			(\mathbf{R}(\mathbf{u}_b)\mathbf{e}_\mathrm{v}) \cdot \mathbf{z}_k & (\mathbf{R}(\mathbf{u}_b)\mathbf{e}_\mathrm{h}) \cdot \mathbf{z}_k \\
			(\mathbf{R}(\mathbf{u}_b)\mathbf{e}_\mathrm{v}) \cdot \bar{\mathbf{z}}_k & (\mathbf{R}(\mathbf{u}_b)\mathbf{e}_\mathrm{h}) \cdot \bar{\mathbf{z}}_k
		\end{bmatrix}\in \mathbb{C}^{2\times 2},
	\end{align}
where $\mathbf{R}(\mathbf{u}_b)\mathbf{e}_\mathrm{v}$ and $\mathbf{R}(\mathbf{u}_b)\mathbf{e}_\mathrm{h}$ represent the mapping of the local CCS of the \(\mathcal{V}\)-element/\(\mathcal{H}\)-element of the PA subarray at the BS to the global CCS.
Similarly, the receive field components are obtained by projecting the LoS signal direction onto the receive antenna using the projection matrix
\begin{align}
	\mathbf{Q}_k (\mathbf{u}_{k}^{\mathrm{r}})= 
	\begin{bmatrix}
		\mathbf{z}_k \cdot (\mathbf{R}(\mathbf{u}_{k}^{\mathrm{r}})\mathbf{e}_\mathrm{v}) & \bar{\mathbf{z}}_k \cdot (\mathbf{R}(\mathbf{u}_{k}^{\mathrm{r}})\mathbf{e}_\mathrm{v}) \\
		\mathbf{z}_k \cdot (\mathbf{R}(\mathbf{u}_{k}^{\mathrm{r}})\mathbf{e}_\mathrm{h}) & \bar{\mathbf{z}}_k \cdot (\mathbf{R}(\mathbf{u}_{k}^{\mathrm{r}})\mathbf{e}_\mathrm{h})
	\end{bmatrix}\in \mathbb{C}^{2\times 2},
\end{align}
where $\mathbf{R}(\mathbf{u}_{k}^{\mathrm{r}})\mathbf{e}_\mathrm{v}$ and $\mathbf{R}(\mathbf{u}_{k}^{\mathrm{r}})\mathbf{e}_\mathrm{h}$ denote the transformation of the local CCS of the \(\mathcal{V}\)-element/\(\mathcal{H}\)-element of the user's PA to the global CCS.
Consequently, the dual-polarized response matrix between the $k$-th user and the dual-polarized antennas on the \( b \)-th PA subarray at the BS is given by  
\begin{align}\label{po3}
	{\mathbf{A}_{k,b}(\mathbf{u}_b,\mathbf{u}_{k}^{\mathrm{r}})} = \mathbf{Q}_k (\mathbf{u}_{k}^{\mathrm{r}})\mathbf{P}_{k,b}(\mathbf{u}_b)\in \mathbb{C}^{2\times 2}.
\end{align}

When the transmission distance is sufficiently large, the phase variation due to the LoS channel is the same for both \(\mathcal{V}\)- and \(\mathcal{H}\)-ports
of the antenna, regardless of the pair of transmit and receive antennas. Therefore, the unpolarformed channel, $\overline{\mathbf{h}}_{k,b}(\mathbf{q}_b,\mathbf{u}_b)$, from the \(\mathcal{V}\)- and
\(\mathcal{H}\)-ports of user $k$ to those of the $b$-th PA subarray at the BS can be expressed as follows, using unpolarformed LoS channel $\mathbf{h}_{k,b}^{\mathrm{LoS}}(\mathbf{q}_{b},\mathbf{u}_{b})$ and dual-polarized response matrix $\mathbf{A}_{k,b}(\mathbf{u}_b,\mathbf{u}_{k}^{\mathrm{r}})$:
\begin{align}\label{tu5}
\!\! \overline{\mathbf{h}}_{k,b}(\mathbf{q}_b,\mathbf{u}_b) = \mathbf{h}_{k,b}^{\mathrm{LoS}}(\mathbf{q}_b,\mathbf{u}_b) \otimes \mathbf{A}_{k,b}(\mathbf{u}_b,\mathbf{u}_{k}^{\mathrm{r}})\in \mathbb{C}^{2N\times 2},
\end{align}
which captures the effects of wireless signal propagation across all potential polarization states.
Different from the dual-polarized antenna \cite{duala}, which requires two radio frequency (RF) chains since each antenna has two ports, each PA is connected to a single RF chain for subsequent signal processing, thereby reducing power consumption (see Fig. \ref{practical_scenario}). 
The $N$- dimensional {\emph{PA polarformed channel}} is then obtained by incorporating the polarization effect for each antenna through  receive polarforming vector $\mathbf{w}_k$ and transmit polarforming vector $\mathbf{v}_b$, which is expressed as
\begin{align}
&\mathbf{h}_{k,b}(\mathbf{q}_b,\mathbf{u}_b,\mathbf{w}_k,\mathbf{v}_b) =(\mathbf{I}_N\otimes\mathbf{v}_b^H)\overline{\mathbf{h}}_{k,b}
(\mathbf{q}_b,\mathbf{u}_b)\mathbf{w}_k
 \label{pcc}\\
&=\underbrace{\mathbf{h}_{k,b}^{\mathrm{LoS}}(\mathbf{q}_b,\mathbf{u}_b)}_{\text{Stable unpolarformed channel}}\times~ 
	\underbrace{\left(\mathbf{v}_b^H\mathbf{A}_{k,b}(\mathbf{u}_b,\mathbf{u}_{k}^{\mathrm{r}})\mathbf{w}_k\right)}_{\text{Dynamic polarformed channel}},\label{pcc1}
\end{align}
where \(\mathbf{I}_N \otimes \mathbf{v}_b^H\) in \eqref{pcc} is due to the fact that all \(N\) antennas in the same PA subarray at BS use the same transmit polarforming vector \(\mathbf{v}_b\).
The polarforming vectors effectively combine the entries of channel $\overline{\mathbf{h}}_{k,b}(\mathbf{q}_b,\mathbf{u}_b)$ according to the polarization of each
antenna at the transmitter/receiver. 

\textbf{Remark 1:} The channel decomposition given in \eqref{pcc1}, introduced in this study for the first time, offers the advantage of decoupling the stable unpolarformed LoS channel component from the dynamic polarformed channel component due to transmitter/receiver polarforming and user rotations in 3D space. In this way, the unpolarformed channel remains unaffected when varying the polarforming vectors and user rotations. This distinctive PA-specific channel representation can be utilized to design efficient localization algorithms that leverage the common parameters in the stable unpolarformed channel across different controllable polarforming vectors  (see Section IV). Furthermore, the elegant structure of \eqref{pcc1} facilitates the optimization of polarforming at both the transmitter and receiver that will be elaborated in Section V.
 
\section{Two-Timescale Protocol Design for PA-Enhanced ISAC}
 \begin{figure}[t]
 	\vspace{-0.45cm}
	\centering
	\setlength{\abovecaptionskip}{0.cm}
	\includegraphics[width=3.39in]{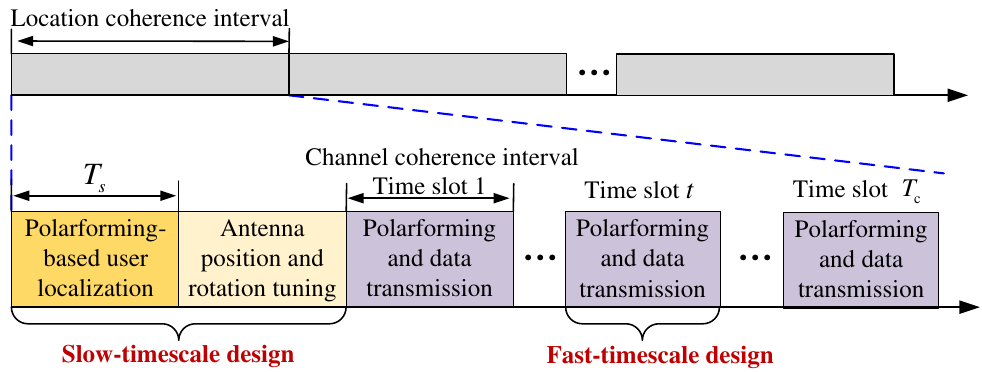}
	\caption{Illustration of the proposed protocol for PA-enhanced ISAC system.}
	\label{protocol}
\vspace{-0.55cm}
\end{figure}
In this section, we propose a two-timescale transmission protocol for the PA-enhanced ISAC system, as illustrated in Fig. \ref{protocol}. Based on our proposed channel setup with users' quasi-static locations and fast-varying rotations, we divide the user transmissions into multiple user location coherence time intervals, during each of which users' locations are assumed to be constant. Each location coherence interval is further divided into two phases, namely, the slow-timescale phase for user localization and BS antenna position/rotation tuning, and the fast-timescale phase for polarforming adapting to users' instantaneous orientations and their induced channel variations.
Specifically, the detailed operation and design in these two phases are provided as follows.   
\begin{itemize}
\item{Phase I (Slow Timescale Design) :}
During this initial phase, user locations are first sensed, and then BS antennas' positions and rotations are determined based on the sensed user locations. To this end, the BS receives the pilot signals from the user devices which are transmitted using time-varying user polarforming vectors \(\mathbf{w}_k\) to enable polarforming-based user localization at the BS (see Section IV). Then, based on the sensed user locations, we optimize the positions and rotations of all PAs at the BS for maximizing the users' average achievable rates (see Section V). The optimized antenna positions and rotations will then be implemented at the BS and remain unchanged during Phase II of each location coherence interval.

\item{Phase II (Fast Timescale Design) :}
The remaining time of each location coherence interval consists of $T_{\mathrm{c}}$ channel coherence intervals. The users' rotations (or induced channels) remain unchanged in each channel coherence interval. During each channel coherence interval, the transmit and receive polarforming vectors are jointly determined based on the instantaneous users' rotations/channels  
to maximize the achievable rates of users. The user locations sensed in Phase I can help to obtain the instantaneous CSI in each channel coherence interval more efficiently, as the user locations are constant with only the user orientations varying over different channel coherence intervals \footnote{The user's location may also vary slightly during each location coherence interval. However, for our considered channel setup assuming far-field LoS propagation for all users, small changes in the user's position do not affect the effective channel gain.}. 
\end{itemize}	

Note that by adapting to user rapid rotation with polarforming after antenna positions and rotations are set to their optimized values based on user locations, our scheme preserves the performance benefits of adapting to both the random and constant components of the PA  channel, thus achieving a good balance between performance and implementation cost.
\vspace{-5pt}
\section{Polarforming-based Localization}
In our proposed PA-enhanced ISAC system, the accuracy of user localization significantly impacts the performance of polarforming and antenna position/rotation optimization. Once user location is determined, the stable unpolarformed channel component given in \eqref{pcc1} between the user and all potential PA positions and rotations at the BS can be reconstructed for slow-timescale antenna position and rotation optimization. Additionally, the user location-related dynamic polarformed channel component in \eqref{pcc1} can be more efficiently acquired for fast-timescale polarforming optimization. Acquiring the stable unpolarformed channel and the dynamic polarformed channel components separately
is a practically appealing approach for user localization, as these separate channels
provide different information about user locations. Therefore, this section presents a polarforming-based user localization algorithm, where the user polarforming vectors dynamically vary according to a designed sequence to help extract the stable unpolarformed LoS channel components. The user locations are then derived based on the decoupled LoS channel components.
\subsection{Extraction of Unpolarformed Channel Component}
To enable user localization sensing with a small number of BS antenna position-rotation placement pairs, PA subarrays at the BS move across a set of $M$ different position-rotation pairs to collect pilot signals from all users. It is generally not feasible to sense the locations of all users widely distributed in the network using a single PA subarray placed at a fixed position-rotation pair at the BS. This is due to the directional sparsity of PA channels \cite{6dmac}, where each user exhibits significant channel gains only with a small subset of PA position-rotation pairs at the BS. 

\begin{figure}[t]
 		\vspace{-0.5cm}
	\centering
	\setlength{\abovecaptionskip}{0.cm}
	\includegraphics[width=2.5in]{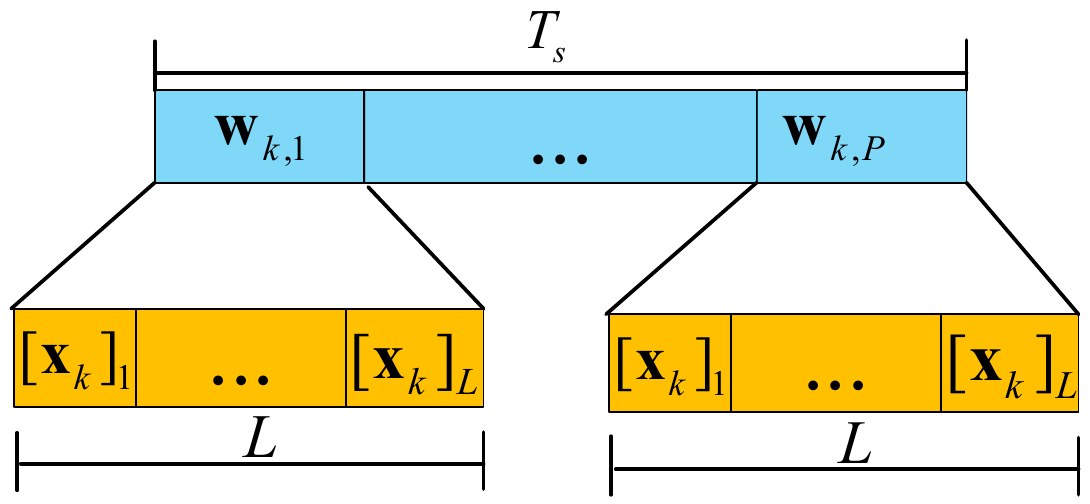}
	\caption{Structured pilot-polarforming pattern in the time domain.}
	\label{pilot}
\vspace{-0.71cm}
\end{figure}

To determine the user locations, we propose a random configuration of polarforming vectors \( \mathbf{w}_{k,p}\) for user \(k\), $p=1, \cdots, P$ with $P$ denoting the total number of pilot signal blocks used for user localization. BS polarforming vectors \( \mathbf{v}_m \), \( m = 1,2,\dots,M \), are set to fixed values for sensing. As shown in Fig. \ref{pilot}, we assume that all user channels are constant during the location training duration (normalized to the symbol period), denoted by $T_{\mathrm{s}}$ in Phase I of the proposed protocol (see Fig. \ref{protocol}), where $T_{\mathrm{s}} = PL$ with $L$ denoting the number of time slots in each pilot signal block. Each user transmits known pilot signals \( \mathbf{x}_k \in \mathbb{C}^{L\times 1} \) to the BS, and the pilot is repeated over the \( P \) blocks. The structured pilot-polarforming pattern is shown in Fig. \ref{pilot}, where 
user polarforming vector \( \mathbf{w}_{k,p}\) remains constant over the $L$ time slots of the $p$-th block and varies from block to block. 

Under the proposed pilot-polarforming pattern, the received signal at the BS, denoted by \(\mathbf{Y}_m \in \mathbb{C}^{L \times N} \), \( m \in\mathcal{M}=\{1,\cdots, M\} \), for the \( m \)-th training position-rotation pair can be expressed as
\begin{align}
	\mathbf{Y}_{m,p}&=\sum_{k=1}^K\eta_{m,k,p}\mathbf{x}_k  \mathbf{h}_{k,m}^{\mathrm{LoS}}(\mathbf{q}_m,\mathbf{u}_m)^T+
	\mathbf{W}_{m,p}\label{aps}\\
	&
	=\mathbf{X}\text{diag}([\boldsymbol{\Omega}_m]_{p,:})\mathbf{H}_m+\mathbf{W}_{m,p}, \label{bbm}
\end{align}
where 
\( [\boldsymbol{\Omega}_m]_{p,:} \) represents the \( p \)-th row of the complex-valued matrix \( \boldsymbol{\Omega}_m=[\boldsymbol{\eta}_{m,1},\boldsymbol{\eta}_{m,2},\cdots,\boldsymbol{\eta}_{m,P}]^T \in \mathbb{C}^{P \times K}\). Here, the vector of dynamic polarformed channel components is represented as 
\begin{align}
	\boldsymbol{\eta}_{m,p} = [\eta_{m,1,p}, \eta_{m,2,p}, \dots, \eta_{m,K,p}]^T\in \mathbb{C}^{K \times 1},
\end{align}
with $	\eta_{m,k,p} = \mathbf{v}_m^H \mathbf{A}_{m,k}(\mathbf{u}_{m}, \mathbf{u}_{k}^{\mathrm{r}}) \mathbf{w}_{k,p}$ as given in \eqref{pcc1}, for user \(k\) at the \(m\)-th training position-rotation pair for PA subarrays during pilot block \(p\), with \(p \in \{1,2,\dots,P\}\).
In addition, $\mathbf{X}=[\mathbf{x}_1,\cdots,\mathbf{x}_K]\in \mathbb{C}^{L \times K}$ denotes the horizontal stack of all pilots from all users, 
$\mathbf{H}_m=[\mathbf{h}_{m,1}^{\mathrm{LoS}}(\mathbf{q}_m,\mathbf{u}_m),\mathbf{h}_{m,2}^{\mathrm{LoS}}(\mathbf{q}_m,\mathbf{u}_m),\cdots,\mathbf{h}_{m,K}^{\mathrm{LoS}}(\mathbf{q}_m,\mathbf{u}_m)]^T\in \mathbb{C}^{K\times N}, m \in\mathcal{M}$,
denotes the collective unpolarformed LoS channel from all $K$ users to all antennas at the $m$-th PA training  position-rotation pair, and $\mathbf{W}_{m,p} \in \mathbb{C}^{L \times N}$ is the complex-valued additive white Gaussian noise (AWGN) matrix with independent and identically distributed (i.i.d.) entries following the distribution of $\mathcal{CN}(0,\sigma^2)$.

Based on the sensing signal model given in \eqref{bbm}, we proceed to apply the PARAFAC decomposition technique \cite{43,xinran,limei, 9805460} to sense users' locations.
First, we rewrite \eqref{bbm} as
\begin{align}\label{a1ps}
	\mathbf{Y}_{m,p}
	=\mathbf{Z}_{m,p}+\mathbf{W}_{m,p}, 
\end{align}
where \(\mathbf{Z}_{m,p}= \mathbf{X}\text{diag}([\boldsymbol{\Omega}_m]_{p,:})\mathbf{H}_m\in \mathbb{C}^{L \times N}\) is the noiseless version of the received signal. We show that each \(\!(l, n)\!\)-th entry of \(\mathbf{Z}_{m,p}\) with \(l = 1, 2, \ldots, L\) and \(n =1, 2, \ldots, N\) is given  by
\begin{align}
	[\mathbf{Z}_{m,p}]_{l,n} = \sum_{k=1}^K [\mathbf{X}]_{l,k} [\mathbf{H}_m]_{k,n} [\boldsymbol{\Omega}_m]_{p,k}. \label{5}
\end{align}
We construct the three-way matrix, \(\mathbf{Z}_m \in \mathbb{C}^{L \times N \times P}\), which incorporates all \(P\) matrices \(\mathbf{Z}_{m,p}\) from \eqref{5} along its third dimension. The unfolded representations corresponding to mode-1, mode-2, and mode-3 of \(\mathbf{Z}_m\) can then be expressed as follows, respectively:
\begin{align}
\mathbf{Z}_m^1 &\triangleq \left(\mathbf{H}_m^T \circ \boldsymbol{\Omega}_m\right) \mathbf{X}^T \in \mathbb{C}^{PN \times L}, \label{6} \\
 \mathbf{Z}_m^2 &\triangleq \left(\boldsymbol{\Omega}_m \circ \mathbf{X}\right) \mathbf{H}_m \in \mathbb{C}^{LP \times N}, \label{7} \\
 \mathbf{Z}_m^3 &\triangleq \left(\mathbf{X} \circ \mathbf{H}_m^T\right) \boldsymbol{\Omega}_m^T \in \mathbb{C}^{NL \times P}, \label{8}
\end{align}
where \(\mathbf{Z}_m^1\), \(\mathbf{Z}_m^2\), and \(\mathbf{Z}_m^3\) represent the horizontal, lateral, and frontal slices of \(\mathbf{Z}_m\). The noise-corrupted matrix unfoldings in \eqref{7} and \eqref{8} can be rewritten as
\begin{align}
\mathbf{Y}_m^2 &\triangleq \left(\boldsymbol{\Omega}_m \circ \mathbf{X}\right) \mathbf{H}_m + \mathbf{W}_m^2,\label{yy2}\\
\mathbf{Y}_m^3 &\triangleq \left(\mathbf{X} \circ \mathbf{H}_m^T\right) \boldsymbol{\Omega}_m^T + \mathbf{W}_m^3,\label{yy3}
\end{align}
where $\mathbf{W}_m^2$ and $\mathbf{W}_m^3$ denote the AWGN matrices. 
The above PARAFAC decomposition facilitates the user localization algorithm design, as shown in the following.

Specifically, given that \(\mathbf{X}\) is known at the BS receiver, \(\mathbf{H}_m\) and \(\boldsymbol{\Omega}_m\) are iteratively estimated by alternately minimizing the corresponding offset square function using the unfolded forms in \eqref{yy2} and \eqref{yy3}. Based on \eqref{yy2}, during the \(i\)-th iteration, the estimate of \(\mathbf{H}_m\), denoted as \(\widehat{\mathbf{H}}_m^{(i)}\), is obtained by minimizing the following offset function:
\begin{align}
	J\left(\widehat{\mathbf{H}}_m^{(i)}\right) = \left\| \mathbf{Y}_m^2 - \widehat{\mathbf{A}}^2_{(i-1)} \widehat{\mathbf{H}}_m^{(i)} \right\|_F^2, \label{10}
\end{align}
where $\widehat{\mathbf{A}}^2_{(i-1)} \triangleq \left(\widehat{\boldsymbol{\Omega}}_m^{(i-1)} \circ \mathbf{X}\right) \in \mathbb{C}^{LP \times K}$. The closed-form solution for \eqref{10} is given by
\begin{align}
	\widehat{\mathbf{H}}_m^{(i)} = \left(\widehat{\mathbf{A}}^2_{(i-1)}\right)^{\dagger} \mathbf{Y}_m^2. \label{11}
\end{align}
Similarly, based on  \eqref{yy3}, during the \(i\)-th iteration, the estimate of \(\boldsymbol{\Omega}_m\), denoted as \(\widehat{\boldsymbol{\Omega}}_m^{(i)}\), is obtained by minimizing the following offset square function:  
\begin{align}
	J\left(\widehat{\boldsymbol{\Omega}}_m^{(i)}\right) = \left\| \mathbf{Y}_m^3 - \widehat{\mathbf{A}}^3_{(i)} \widehat{\boldsymbol{\Omega}}_m^{(i)} \right\|_F^2, \label{12}
\end{align}
where $\widehat{\mathbf{A}}^3_{(i)} \triangleq \left(\mathbf{X} \circ (\widehat{\mathbf{H}}_m^{(i)})^T\right)  \in \mathbb{C}^{LN \times K}$. The closed-form solution for \eqref{12} is given by
\begin{align}
	\widehat{\boldsymbol{\Omega}}_m^{(i)} = \left(\widehat{\mathbf{A}}^3_{(i)}\right)^{\dagger} \mathbf{Y}_m^3. \label{13}
\end{align}

Note that we initialize \(\hat{\mathbf{H}}_m^{(0)}\) and \(\hat{\boldsymbol{\Omega}}_m^{(0)}\) as the eigenvector matrices corresponding to the \(K\) non-zero eigenvalues of \(\left(\mathbf{Y}_m^2\right)^{\mathrm{H}} \mathbf{Y}_m^2\) and \(\left(\mathbf{Y}_m^3\right)^{\mathrm{H}} \mathbf{Y}_m^3\), respectively. The proposed iterative alternating algorithm terminates when the normalized mean square error (NMSE) between any two adjacent iterations is less than a given threshold $\kappa$. Note that the iterative estimations in Algorithm 1 may encounter a scaling ambiguity issue, which can be resolved through normalization as in \cite{43}.
\vspace{-3pt}
\subsection{Unpolarformed Channel-Based Localization}
\vspace{-3pt}
Based on the estimated stable unpolarformed channel, \(\widehat{\mathbf{H}}_m\) for \(m=1,\cdots, M\), the users' locations are determined by extracting their corresponding  DoAs at the BS and user-BS distances. By
utilizing the orthogonality of the signal and noise subspaces,
one practical method  for DoA estimation is using the multiple signal classification (MUSIC) algorithm \cite{MUSIC}.

To apply MUSIC, we first construct a sample covariance matrix from the channel estimates. All \(\widehat{\mathbf{H}}_m\) matrices are stacked into a larger matrix
$\hat{\mathbf{H}} = 
	\begin{bmatrix}
		\hat{\mathbf{H}}_1 & \hat{\mathbf{H}}_2 & \cdots & \hat{\mathbf{H}}_M
	\end{bmatrix}^T
	\;\in\; \mathbb{C}^{MN \times K}$ and channel covariance
matrix is calculated as
\begin{align}
	\widehat{\mathbf{R}} &= \frac{1}{K} \hat{\mathbf{H}} \hat{\mathbf{H}}^H \;\in\; \mathbb{C}^{MN \times MN}. \label{eq:R_cov}
\end{align}

Based on \eqref{gen}, the steering vector for $M$ training position-rotation pairs is constructed as
\begin{align}
	\!\!\!\!\!&\mathbf{a}_k(\mathbf{f}_k)\! =\!
	\begin{bmatrix}
		\mathbf{a}_{k,1}^T(\mathbf{q}_1, \mathbf{u}_1; \mathbf{f}_k) \!
		\cdots \!
		\mathbf{a}_{k,M}^T(\mathbf{q}_M, \mathbf{u}_M; \mathbf{f}_k)
	\end{bmatrix}^T. 
\end{align}
Next, we perform the eigenvalue decomposition (EVD) of the covariance matrix \(\widehat{\mathbf{R}}\), expressed as 
\begin{align}
	\widehat{\mathbf{R}} &= \mathbf{U}_s \boldsymbol{\Lambda}_s \mathbf{U}_s^H + \mathbf{U}_n \boldsymbol{\Lambda}_n \mathbf{U}_n^H, \label{eq:eigendecomp}
\end{align}
where $\boldsymbol{\Lambda}_s$ is a diagonal matrix containing the
largest $K$ eigenvalues of \(\widehat{\mathbf{R}}\), and \(\mathbf{U}_s\) contains the eigenvectors corresponding to the \(K\) largest eigenvalues, while the rest of the eigenvalues and eigenvectors, respectively, constitute $\boldsymbol{\Lambda}_n$ and $\mathbf{U}_n$.

The DoAs of all \(K\) users are obtained by finding top-$K$ peaks in the pseudo-spectrum \cite{MUSIC}, i.e., 
\begin{align}
	\hat{\mathbf{f}}_1, \hat{\mathbf{f}}_2, \dots, \hat{\mathbf{f}}_K
	&= \underset{\mathbf{f}_k}{\mathrm{arg top\!\!-\!\!K}}\;
	\left( \frac{1}{\bigl\|\mathbf{U}_n^H \mathbf{a}(\mathbf{f}_k)\bigr\|^2} \right). \label{eq:doa_estimation}
\end{align}

Once the DoA vectors are obtained, the user-BS distance for user \(k\) can be calculated. 
Using the observed channel amplitude  \(\|\hat{\mathbf{h}}_{k,m}^{\mathrm{LoS}}\|\) of the stable unpolarized channel  across all \(M\) LoS channel estimates  and the user-BS geometry, we formulate the following least squares (LS)-based  minimization problem for estimating \(d_k\):
\begin{align} \label{y6}
\!\!\!\!\!\!	\hat{d}_k
\!=\!
\arg\min_{d>0}
\sum_{m=1}^M
\Bigl(
\|\hat{\mathbf{h}}_{k,m}^{\mathrm{LoS}}\|
\;-\;
\sqrt{\frac{\epsilon_0N}{d^2}}
\sqrt{g_{k,m}(\mathbf{u}_{m})}
\Bigr)^2.
\end{align}
The solution to the problem in \eqref{y6} has the following closed-form expression:
\begin{align} \label{ddk}
		\hat{d}_k 
	\;=\;
	\sqrt{\epsilon_0N}
	\;\frac{
		\sum_{m=1}^M g_{k,m}(\mathbf{u}_{m})
	}{
		\sum_{m=1}^M 
		\|\hat{\mathbf{h}}_{k,m}^{\mathrm{LoS}}\|\;\sqrt{g_{k,m}(\mathbf{u}_{m})}
	}.
\end{align}

The proposed polarforming-based localization algorithm is summarized in Algorithm 1. The complexity
of Algorithm 1 is dominated by the involved matrix
inverse computations for obtaining  $\mathbf{H}_m$ and $\boldsymbol{\Omega}_m$, which have the complexity orders of \(\mathcal{O}(K^3+4K^2LP-KLP)\) and \(\mathcal{O}(K^3+4K^2NL-KNL)\), respectively. Thus, the total
computational complexity of Algorithm 1 is \(\mathcal{O}(2K^3+4K^2L(P+N)-KL(P+N))\).
\begin{algorithm}[t!]
	\caption{Proposed Polarforming-based Localization}
	\label{alg:ALS_CE}
	\begin{algorithmic}[1]
		\STATE \textbf{Input:} Feasible polarforming $\{\{\mathbf{w}_{k,p}\}_{k=1}^K\}_{p=1}^P$, $M$ candidate position-rotation pairs at the BS, and threshold $\kappa$.
		\STATE \textbf{Initialization:} $\widehat{\mathbf{H}}_m^{(0)}$, $\widehat{\boldsymbol{\Omega}}_m^{(0)}$, and set the algorithmic iteration as $i = 1$.\\
		\emph{Step I: Extraction of Unpolarformed Channel}:
		\FOR{$m = 1, 2, \ldots, M$}
		\FOR{$i = 1, 2, \ldots, I_{\max}$}
		\STATE Set $\mathbf{A}_{(i-1)}^2 \triangleq \left(\boldsymbol{\Omega}_m^{(i-1)} \circ \mathbf{X}\right) $ and then compute 	$\widehat{\mathbf{H}}_m^{(i)} = \left(\widehat{\mathbf{A}}^2_{(i-1)}\right)^{\dagger} \mathbf{Y}_m^2$ in \eqref{10}.
		\STATE Set $\mathbf{A}_{(i)}^3 \triangleq \left(\mathbf{X} \circ \left(\mathbf{H}_m^{(i)}\right)^T\right)$ and compute $	\widehat{\boldsymbol{\Omega}}_m^{(i)} = \left(\widehat{\mathbf{A}}^3_{(i)}\right)^{\dagger} \mathbf{Y}_m^3$ in \eqref{13}.
		\IF{$\frac{\|\widehat{\mathbf{H}}_m^{(i)} - \widehat{\mathbf{H}}_m^{(i-1)}\|_F^2}{\|\widehat{\mathbf{H}}_m^{(i)}\|_F^2} \leq \kappa$ and $\frac{\|\widehat{\boldsymbol{\Omega}}_m^{(i)} - \widehat{\boldsymbol{\Omega}}_m^{(i-1)}\|_F^2}{\|\widehat{\boldsymbol{\Omega}}_m^{(i)}\|_F^2} \leq \kappa$}
		\STATE \textbf{break}
		\ENDIF
		\ENDFOR
		\ENDFOR
		\STATE Obtain $\widehat{\mathbf{H}}_m^{(i)}$ and $\widehat{\boldsymbol{\Omega}}_m^{(i)}$ that are the estimations of $\mathbf{H}_m$ and ${\boldsymbol{\Omega}}_m$ for $m=1,2,\cdots,M$, respectively.\\
		\emph{Step II: Unpolarformed Channel-Based Localization}:
		\STATE Compute the DoA vectors for all users using \eqref{eq:doa_estimation}.  
		\STATE Compute the distances for all users using \eqref{ddk}.
		\STATE \textbf{Output:} $\hat{\mathbf{f}}_k$ and $d_k$ for all users $k=1,2,\cdots, K$.
	\end{algorithmic}
\end{algorithm}
\vspace{-5pt}
\section{Polarforming, Position, and Rotation Optimization}
\vspace{-3pt}
In this section, we formulate and solve the optimization problem for polarforming design (see Section V-A) in Phase II and antenna position and rotation design (see Section V-B) in Phase I of the proposed protocol.
As shown in Fig. \ref{practical_scenario}, the downlink received signal at the $k$-th user is given by
\begin{align}\label{X1}
	y_k=
	\mathbf{h}_k(\mathbf{q},\mathbf{u},\mathbf{w}_k,\mathbf{v})^H\sum_{j=1}^K
	\mathbf{c}_jx_j+n_k,~k\in \mathcal{K},
\end{align}
where $x_k \sim \mathcal{CN}(0, 1)$ is the data symbol intended for the $k$-th
user, $\mathbf{c}_k\in \mathbb{C}^{NB\times 1}$ represents the transmit precoder for user $k$, ${\mathbf{q}}=[\mathbf{q}_1^T,\cdots,\mathbf{q}_B
^T]^T\in \mathbb{R}^{3B\times 1},\label{lk}$ and 
$\mathbf{u}=[\mathbf{u}_1^T,\mathbf{u}_2^T,\cdots,\mathbf{u}_B^T]^T\in \mathbb{R}^{3B\times 1} \label{lk1}$ are position and rotation vectors for all PA subarrays at BS, and $n_k\sim\mathcal{CN}(0,\sigma^2)$ is the AWGN. Using Shannon’s formula, the achievable rate of the $k$-th user over unit radio spectrum bandwidth is given by
\begin{align}\label{r}
	&{R}_k(\mathbf{q},\mathbf{u},\mathbf{w}_k,\mathbf{v},\mathbf{c})= \nonumber\\
	&\log_2\left(1+
	\frac{|\mathbf{h}_k(\mathbf{q},\mathbf{u},\mathbf{w}_k,\mathbf{v})^H
		\mathbf{c}_k|^2}{\sum\limits_{j\in\mathcal{K}\setminus k}|\mathbf{h}_k(\mathbf{q},\mathbf{u},\mathbf{w}_k,\mathbf{v})^H
		\mathbf{c}_j|^2+\sigma^2}\right), 
\end{align}
where
$\mathbf{h}_k(\mathbf{q},\mathbf{u},\mathbf{w}_k,\mathbf{v})=\left[\mathbf{h}_{k,1}^T,
	\mathbf{h}_{k,2}^T,\cdots,\mathbf{h}_{k,B}^T\right]^T\in \mathbb{C}^{NB\times 1}
$ denotes the overall PA channel between user $k$ and all the PA subarrays at the BS, $\mathbf{v}=[\mathbf{v}_1^T,\cdots,\mathbf{v}_B^T]^T$, and $\mathbf{c}=[\mathbf{c}_1^T,\cdots,\mathbf{c}_K^T]^T$.

Following the proposed two-timescale protocol,
our design objective is to maximize the weighted sum-rate for all users
by jointly optimizing BS-side antenna positions $\mathbf{q}$ and rotations $\mathbf{u}$ in the slow
timescale, as well as BS-side transmit polarforming vectors $\mathbf{v}$ and precoding vectors $\mathbf{c}$ and user-side polarforming vectors $\{\mathbf{w}_k\}$ in the fast timescale. This leads to the following optimization problem: 
\begin{subequations}
	\label{MG3}
	\begin{align}
		\!\text{(P1) :}~\!\!\mathop{\max}\limits_{\mathbf{q},\mathbf{u}}~&~\mathbb{E}\left[
		\mathop{\max}\limits_{\{\mathbf{w}_k\}_{k=1}^K, \mathbf{v},\mathbf{c}}
		\sum\limits_{k\in\mathcal{K}}\varrho_k{R}_k(\mathbf{q},
		\mathbf{u},\mathbf{w}_k,\mathbf{v},\mathbf{c})\right]\\
		\text {s.t.}~&~[\mathbf{w}_k]_i\in \mathcal{F}, \forall k\in \mathcal{K}, i\in \{1,2\}, \label{pc1}\\
		~&~ [\mathbf{v}_b]_i\in \mathcal{F}, \forall b\in \mathcal{B}, i\in \{1,2\}, \label{pc2}\\
		~&~\sum_{k \in \mathcal{K}} \|\mathbf{c}_k\|^2 \leq \zeta, \label{pow1}\\
		~&~\mathbf{q}_i\in\mathcal{C}, ~\forall i \in \mathcal{B},  \label{M1}\\
		~&~ \|\mathbf{q}_i-
		\mathbf{q}_{j}\|\geq d_{\min},~\forall i ,j \in \mathcal{B}, ~j\neq i,\label{M2}\\
		~&~ \mathbf{n}(\mathbf{u}_i)^T(\mathbf{q}_{j}-\mathbf{q}_i)\leq  0,~\forall i ,j \in \mathcal{B}, ~j\neq i, \label{M3}
	\end{align}
\end{subequations}
where $\varrho_k$ represents the rate weight of user $k$, $\zeta$ is the total transmit power of the BS, the expectation is taken over the random channel variations due to arbitrary user rotations, and constraints \eqref{pc1} and \eqref{pc2} ensure that the receive and transmit polarforming vectors satisfy the discrete amplitude and phase requirements. Constraint \eqref{M1} ensures that the center of each PA subarray is positioned within the convex 3D site space of the BS. As discussed in \cite{6dmatwc}, minimum distance \(d_{\min}\) enforced by constraint \eqref{M2} prevents overlapping and coupling between PA subarrays. Constraint \eqref{M3} mitigates mutual signal reflections among BS antennas.
\vspace{-5pt}
\subsection{Fast Timescale Optimization}
\vspace{-3pt}
During each channel coherence interval, the BS first estimates the instantaneous channels of all users,  $\mathbf{h}_k(\mathbf{q},\mathbf{u},\mathbf{w}_k,\mathbf{v})$, for all possible $\{\mathbf{w}_k\}$ and $\mathbf{v}$ and with fixed antenna position vector $\mathbf{q}$, rotation vector $\mathbf{u}$, and user locations. Then, the BS determines its polarforming $\mathbf{v}$, precoding $\mathbf{c}$ as well as user polarforming $\{\mathbf{w}_k\}$. Given $\mathbf{q}$ and $\mathbf{u}$, problem (P1) is simplified to 
\begin{subequations}
	\label{op2}
	\begin{align}
		\text{(P2) :}~&~
		\mathop{\max}\limits_{\{\mathbf{w}_k\}_{k=1}^K, \mathbf{v}, \mathbf{c}}
		\sum\limits_{k\in\mathcal{K}}\varrho_k{R}_k(\mathbf{q},
		\mathbf{u},\mathbf{w}_k,\mathbf{v},\mathbf{c})\\
		~&~\text {s.t.}~\eqref{pc1},\eqref{pc2},\eqref{pow1}.
	\end{align}
\end{subequations}

To reformulate problem (P2) into a more tractable form, we employ the weighted minimum mean
squared error (WMMSE) method \cite{38}.
Assuming signal \( x_k \) is decoded using equalizer \( \xi_k \), the estimated signal at user \( k \) is given by 
\( \hat{x}_k = \xi_k y_k\). The mean square error (MSE) for user \( k \), defined as \( e_k = \mathbb{E}[ |\hat{x}_k - x_k|^2 ] \), is derived as  
\begin{align}
	e_k& = |\xi_k|^2 \bigg( \sum_{j\in\mathcal{K} }|\mathbf{h}_k(\mathbf{q},\mathbf{u},\mathbf{w}_k,\mathbf{v})^H
	\mathbf{c}_j|^2+\sigma^2 \bigg)\nonumber\\
	&
	- 2 \operatorname{Re} \big\{ \xi_k^* \mathbf{h}_k(\mathbf{q},\mathbf{u},\mathbf{w}_k,\mathbf{v})^H
	\mathbf{c}_k \big\} + 1.
	\label{eq:MSE_expression}
\end{align}

Then, problem (P2) shares the same globally optimal solution with the following WMMSE problem:
\begin{subequations}
	\label{op2-1}
	\begin{align}
		\text{(P2-1) :}~&~
		\mathop{\min}\limits_{\{\mathbf{w}_k,\xi_k,\epsilon_k\}_{k=1}^K, \mathbf{v},\mathbf{c}}~~\sum\limits_{k\in\mathcal{K}}\varrho_k(\epsilon_ke_k-\log_2(\epsilon_k))
		\\
		~&~\text {s.t.}~\eqref{pc1},\eqref{pc2}, \eqref{pow1},
	\end{align}
\end{subequations}
where $\epsilon_k$ denotes the weighting factor for user $k$.

To facilitate parallel and element-wise updates of the elements in polarforming vectors $\{\mathbf{w}_k\}$ and $\{\mathbf{v}_b\}$, thereby simplifying their optimization,
we introduce auxiliary variables $\{\overline{\mathbf{w}}_k\}$ and $\{\overline{\mathbf{v}}_b\}$. As a result, problem (P2-1) is equivalently transformed to
\begin{subequations}
	\label{aop2-1}
	\begin{align}
		\text{(P2-2) :}&
		\mathop{\min}\limits_{\{\mathbf{w}_k,\xi_k,\epsilon_k\}_{k=1}^K, \{\mathbf{v}_b\}, \mathbf{c}}~\sum\limits_{k\in\mathcal{K}}\varrho_k(\epsilon_ke_k-\log_2(\epsilon_k))
		\\
		\text {s.t.}&~\sum_{k \in \mathcal{K}} \|\mathbf{c}_k\|^2 \leq \zeta, \label{3pow1}\\
				~&~\mathbf{w}_k=\overline{\mathbf{w}}_k, k\in\mathcal{K}, \label{0pcc1}\\
			~&~\mathbf{v}_b=\overline{\mathbf{v}}_b, b\in\mathcal{B},\label{0pcc01}\\
			~&~[\overline{\mathbf{w}}_k]_i\in \mathcal{F}, \forall k\in \mathcal{K}, i\in \{1,2\}, \label{3pc1}\\
			~&~ [\overline{\mathbf{v}}_b]_i\in \mathcal{F}, \forall b\in \mathcal{B}, i\in \{1,2\}. \label{3pc2}
	\end{align}
\end{subequations}

Next, we employ the penalty dual decomposition (PDD) framework to develop a double-loop iterative algorithm for solving (P2-2). The inner loop addresses an augmented Lagrangian problem using block-based minimization, while the outer loop updates the dual variables and penalty coefficients based on constraint violations until convergence. 

Specifically, in the inner loop of PDD, we apply the block coordinate descent (BCD) method to solve the following augmented Lagrangian problem of (P2-2):
\begin{subequations}
	\label{aop33-1}
	\begin{align}
	\text{(P2-3) :}&
		\mathop{\min}\limits_{\{\mathbf{w}_k,\xi_k,\epsilon_k\}_{k=1}^K, \{\mathbf{v}_b\},\mathbf{c}}~~\sum\limits_{k\in\mathcal{K}}\varrho_k(\epsilon_ke_k-\log_2(\epsilon_k))+\nonumber\\
		&\!\!\!\frac{1}{2 \mu }\sum_{k\in\mathcal{K}}\|\mathbf{w}_k-\overline{\mathbf{w}}_k+ \mu\mathbf{t}_k \|^2	\!+\!\frac{1}{2\mu}\sum\limits_{b\in\mathcal{B}} \left\| \mathbf{v}_b - \overline{\mathbf{v}}_b + \mu \bar{\mathbf{t}}_b \right\|^2,\nonumber
		\\
		\text {s.t.}&~\eqref{3pow1}, \eqref{3pc1},\eqref{3pc2}, 
	\end{align}
\end{subequations}
where \(\mathbf{t}_k\) and \(\bar{\mathbf{t}}_b\) represent the dual variable vectors corresponding to  constraints \(\mathbf{w}_k=\overline{\mathbf{w}}_k\) and \(\mathbf{v}_b=\overline{\mathbf{v}}_b\), respectively, while \(\mu\) is the penalty coefficient.  By dividing the optimization variables into the following blocks: \(\{\mathbf{w}_k\}\), \(\{\bar{\mathbf{w}}_k\}\), \(\{\mathbf{v}_b\}\), \(\{\bar{\mathbf{v}}_b\}\), \(\{\xi_k\}\), \(\{\epsilon_k\}\), and \(\{\mathbf{c}_k\}\),  each block can be optimized separately while keeping the others fixed. 

First, the user polarforming vectors \(\{\mathbf{w}_k\}\) are updated by solving the following unconstrained quadratic program (QP) problem:
	\label{op25}
	\begin{align}
		\!\text{(P2-3.1) :}~&
		\mathop{\min}\limits_{\{\mathbf{w}_k\}}~~\sum\limits_{k\in\mathcal{K}}\varrho_k\epsilon_k |\xi_k|^2 
		\sum_{j\in\mathcal{K}}
		\bigl|
		\mathbf{w}_k^H
		\mathbf{M}_{k}^H
		\mathbf{c}_j
		\bigr|^2
		-\sum_{k\in\mathcal{K}}2\varrho_k\epsilon_k\nonumber\\
		&\!\!\!\!\!\!\!\!\!\!\!\!\!\!\!
		\mathrm{Re}\Bigl\{
		\xi_k^*
		\mathbf{w}_k^H
		\mathbf{M}_{k}^H
		\mathbf{c}_k
		\Bigr\}+\frac{1}{2 \mu }\sum_{k\in\mathcal{K}}\|\mathbf{w}_k-\overline{\mathbf{w}}_k+ \mu\mathbf{t}_k \|^2,
	\end{align}
which is derived by substituting the relationship $\mathbf{h}_k(\mathbf{q}, \mathbf{u}, \mathbf{w}_k, \mathbf{v})
=
\mathbf{M}_k \mathbf{w}_k$ into problem (P2-3) and neglecting the terms that do not involve 
$\{\mathbf{w}_k\}$, where
$\mathbf{M}_k =
\big[\mathbf{h}_{k,1}^{\mathrm{LoS}} \bigl(\mathbf{v}_1^H \mathbf{A}_{k,1}\bigr)^T, 
\cdots, 
\mathbf{h}_{k,B}^{\mathrm{LoS}} \bigl(\mathbf{v}_B^H \mathbf{A}_{k,B}\bigr)^T
\big]^T
\in \mathbb{C}^{NB \times 2}$.
The closed-form optimal solution of (P2-3.1) can be expressed as
\begin{align}\label{50}
	\mathbf{w}_k^{\mathrm{opt}}
	=
	\mathbf{C}_k^{-1}\mathbf{b}_k,
\end{align}
where
\begin{align}
	\mathbf{C}_k 
	&=
	2\varrho_k\epsilon_k|\xi_k|^2\sum_{j\in\mathcal{K}} \mathbf{M}_k^H\mathbf{c}_j\mathbf{c}_j^H\mathbf{M}_k 
	+ \tfrac{1}{\mu}\mathbf{I}, \\
	\mathbf{b}_k 
	&= 
	2\varrho_k\epsilon_k\xi_k\mathbf{M}_k^H\mathbf{c}_k
	+ \tfrac{1}{\mu}\Big(\overline{\mathbf{w}}_k - \mu\mathbf{t}_k\Big).
\end{align}

Next, the $\overline{\mathbf{w}}_k$-subproblem is
given by  
\begin{subequations}
	\label{ppp}
	\begin{align}
		\text{(P2-3.2) :} ~\min_{\overline{\mathbf{w}}_k}~&~\|\mathbf{w}_k - \overline{\mathbf{w}}_k + \mu \mathbf{t}_k \|^2  \\
		\text {s.t.}~&~[\overline{\mathbf{w}}_k]_i \in \mathcal{F}, \quad \forall i \in \{1, 2\}.
	\end{align}
\end{subequations}
Since the elements of \(\overline{\mathbf{w}}_k\) are independent in both the objective function and constraints, the optimal solution can be computed in parallel as
\begin{align}\label{54}
	[\overline{\mathbf{w}}_k]_i^{\text{opt}} = \hat{\rho}_{k,i} e^{j \angle [\overline{\mathbf{w}}_k]_i},
\end{align}
where 
\begin{align} \label{p023}
	&	\angle [\overline{\mathbf{w}}_k]_i = \arg \min_{\angle [\overline{\mathbf{w}}_k]_i \in \mathcal{S}} |\angle [\overline{\mathbf{w}}_k]_i - \angle ([\mathbf{w}_k]_i+\mu [\mathbf{t}_k]_i )|,\\
	& \hat{\rho}_{k,i} = \arg \min_{{\rho}_{k,i} \in \mathcal{A}} |{\rho}_{k,i} e^{j \angle [\overline{\mathbf{w}}_k]_i} - ([\mathbf{w}_k]_i+\mu [\mathbf{t}_k]_i )|.
\end{align}
The detailed solutions for the remaining variables in (P2-3) are provided in Appendix A.


In the outer loop of PDD framework, the dual variables are updated as
\begin{align}
	&\mathbf{t}_k \gets  \mathbf{t}_k + \frac{1}{\mu} (\mathbf{w}_k - \mathbf{t}_k), k\in \mathcal{K}, \label{dual}\\
	&\bar{\mathbf{t}}_b \gets  \bar{\mathbf{t}}_b + \frac{1}{\mu} (\mathbf{v}_b - \overline{\mathbf{v}}_b), b\in \mathcal{B}. \label{dual1}
\end{align}

The proposed polarforming optimization algorithm is summarized in Algorithm 2, which is guaranteed to converge \cite{pdd}. The complexity of Algorithm 2 is \(\mathcal{O}\left(I_{\mathrm{out}} I_{\mathrm{in}} \left( K N^2B^2 \right) \right)\), where \(I_{\mathrm{out}}\) and \(I_{\mathrm{in}}\) respectively denote the outer and inner iteration numbers required for convergence.
\begin{algorithm}[t!]
	\caption{Proposed Polarforming Optimization Algorithm for Solving Problem (P2)}
	\label{alg:PDD}
	\begin{algorithmic}[1]
	\STATE \textbf{Input:} $\mathbf{q}$, $\mathbf{u}$, and $\zeta$.
	\STATE Initialize $\{\mathbf{w}_k\}$, $\{\mathbf{v}_b\}$, and $\{\mathbf{c}_k\}$, set the outer iteration index $i_{\text{out}} = 0$, $\epsilon_{\text{in}} > 0$, $\epsilon_{\text{out}} > 0$, and $ \varpi < 1$.
		\REPEAT
		\STATE Set the inner iteration index $i_{\text{in}} = 0$.
		\REPEAT
		\STATE Update \(\{\mathbf{w}_k\}\), \(\{\bar{\mathbf{w}}_k\}\), \(\{\mathbf{v}_b\}\), \(\{\bar{\mathbf{v}}_b\}\), \(\{\xi_k\}\), \(\{\epsilon_k\}\), and \(\{\mathbf{c}_k\}\) successively according to \eqref{50}, \eqref{54}, and Appendix A.
		\STATE Update the inner iteration index: $i_{\text{in}} \gets i_{\text{in}} + 1$.
		\UNTIL The relative reduction in the objective function of (P2-3) falls below the threshold $\epsilon_{\text{in}}$.
		\STATE Update the dual variables by \eqref{dual} and \eqref{dual1} and update $\mu \gets \varpi\mu$.
		\STATE $i_{\text{out}} \gets i_{\text{out}} + 1$.
		\UNTIL Both constraint violations, $\|\mathbf{v} - \bar{\mathbf{v}}\|_\infty$ and $\|\mathbf{w} - \bar{\mathbf{w}}\|_\infty$, are below the threshold $\epsilon_{\text{out}}$, where $\bar{\mathbf{v}}$ and $\bar{\mathbf{w}}$ represent the collections of $\{\bar{\mathbf{v}}_b\}$ and $\{\bar{\mathbf{w}}_k\}$, respectively.
		\STATE \textbf{Output:} $\{\mathbf{c}_k\}$,  $\{\mathbf{w}_k\}$, and $\{\mathbf{v}_b\}$.
	\end{algorithmic}
\end{algorithm}
\vspace{-5pt}
\subsection{Slow Timescale Optimization}
\vspace{-3pt}
The slow-timescale optimization of antenna position vector $\mathbf{q}$ and rotation vector $\mathbf{u}$ in problem (P1) is a stochastic optimization problem due to the expectation operation in the objective function. Since this objective function is intractable, we approximate it as a deterministic function. Specifically, we independently generate  $\overline{L}$ sets of random channel samples for all users and use their achievable rates averaged over these channel samples as a close approximation of the expected rate in the objective function of (P1), provided that \(\overline{L}\) is sufficiently large. Given the user locations estimated at the beginning of Phase I of our proposed protocol, the BS can obtain the users' LoS channels for all possible PA position-rotation pairs at the BS and accordingly generate the \(\overline{L}\) random channel samples by independently and randomly varying rotations $\mathbf{u}_{k}^{\mathrm{r}}$ of all users.   

Suppose that the $l$-th sample of the PA time-variant channels is $\hat{\mathcal{H}}^l=\{\mathbf{h}_1^l,\mathbf{h}_2^l,\cdots,\mathbf{h}_K^l\}$. The average sum rate of the $k$-th user is approximated by
$\frac{1}{\overline{L}}  \sum_{l=1}^{\overline{L}} R_k\left(\mathbf{q},
	\mathbf{u}, \mathbf{w}_k,\mathbf{v},\mathbf{c} ,\hat{\mathcal{H}}^l\right)$,
where \(\hat{\mathcal{H}}\) represents the collection of all $\overline{L}$ channel samples. Problem (P1) is thus recast as 
\begin{subequations}
	\label{lll}
	\begin{align}
		\text{(P3) :}~\mathop{\max}\limits_{\mathbf{q},\mathbf{u}}~&~
		\sum\limits_{k\in\mathcal{K}}\varrho_k\frac{1}{\overline{L}}  \sum_{l=1}^{\overline{L}} R_k\left(\mathbf{q},
		\mathbf{u}, \mathbf{w}_k,\mathbf{v},\mathbf{c} ,\hat{\mathcal{H}}^l\right) \\
		\text {s.t.}~&~\eqref{M1}, \eqref{M2},\eqref{M3}.
	\end{align}
\end{subequations}

Note that problem (P3) is a non-convex optimization problem due to the non-concave objective function and the non-convex constraints in \eqref{M2} and \eqref{M3}. Traditional convex optimization methods are not practically efficient for solving problem (P3). Inspired by the low complexity of particle swarm optimization (PSO) algorithms \cite{con}, we propose a recursive sampling-based PSO (RS-PSO) scheme for optimizing the position and rotation parameters. Specifically, we assume that \( S \) particles are used to explore the constrained search space, with each particle assigned a velocity at every iteration. The position of a particle, represented as  
\begin{align}
	\mathbf{s}=[\mathbf{q}^T, \mathbf{u}^T]^T\in\mathbb{R}^{6B\times 1},
\end{align}  
includes the optimized antenna position and rotation parameters. The velocity of a particle is given by \( \mathbf{m}\in\mathbb{R}^{6B\times 1} \). 
\begin{algorithm}[t!]
	\caption{Proposed Two-Timescale Polarforming, Position, and Rotation (TT-PPR) Optimization Algorithm}
	\begin{algorithmic}[1]
		\STATE $\textbf{Input:}$ ~$\overline{L}, B, N, K, S$ and 
		$I_{\text{iter}}$.
		\STATE {\textbf{Step 1}}: (Slow timescale optimization for solving (P3))
		\STATE Initialize $\{\mathbf{s}_j^{(0)}\}_{j=1}^S$, $\{\mathbf{m}_j^{(0)}\}_{j=1}^S$ and $\{S,\omega, c_1, c_2, \tau_1, \tau_2, I_{\text{iter}}\}$.
		\STATE Generate \( \overline{L} \) channel samples based on estimated user locations and divide them into \( \overline{N} \) mini-batches.
		\STATE Calculate \eqref{20} using the first mini-batch of samples, where $\{\mathbf{w}_{k,j}^{(0)}, \mathbf{c}_{k,j}^{(0)}\}$ and $\{\mathbf{v}_{b,j}^{(0)}\}$ are obtained by Algorithm 2.
		\STATE Obtain $\{\hat{\mathbf{s}}_j^{(0)}\}_{j=1}^S$, and find the global optimal position:
		\[
		\mathbf{s}_{\mathrm{g}} = \arg \max_{\mathbf{s}} \{J^{(0)}(\hat{\mathbf{s}}_1), J^{(0)}(\hat{\mathbf{s}}_2), \dots, J^{(0)}(\hat{\mathbf{s}}_S)\}.
		\]
		\FOR{$i = 1 : I_{\text{iter}}$}
		\STATE Update particle velocity $\{\mathbf{m}_j^{(i)}\}_{j=1}^S$ and position $\{\mathbf{s}_j^{(i)}\}_{j=1}^S$ by using \eqref{gb} and \eqref{vb}.
		\FOR{$j = 1 : S$}
		\STATE Given particle $\mathbf{s}_j^{(i)}$, calculate the optimal $\{\mathbf{w}_{k,j}^{(i)}\}$, $\{\mathbf{v}_{b,j}^{(i)}\}$, $\{\mathbf{c}_{k,j}^{(i)}\}$ according to Algorithm 2.
		\STATE Evaluate the fitness value of particle $\mathbf{s}_j^{(i)}$ over the $i$th mini-batch of samples by using \eqref{20}.
		\IF{$J^{(i)} > J^{(i-1)}$}
		\STATE $\hat{\mathbf{s}}_j = \mathbf{s}_j^{(i)}$.
		\ELSE
		\STATE $\hat{\mathbf{s}}_j = \mathbf{s}_j^{(i-1)}$.
		\ENDIF
		\ENDFOR
		\STATE Update $
		\mathbf{s}_{\mathrm{g}}= \arg \max_{\hat{\mathbf{s}}} \{J^{(i)}(\hat{\mathbf{s}}_1),  \dots, J^{(i)}(\hat{\mathbf{s}}_S)\}$.
		\ENDFOR
		\STATE $\textbf{Output:} [\mathbf{q}^T, \mathbf{u}^T]^T=\mathbf{s}_{\mathrm{g}}$.
		\STATE {\textbf{Step 2}}: (Fast timescale optimization over $t\in [1, T_{\mathrm{c}}]$)
		\STATE Apply Algorithm 2 given $\mathbf{q}$, $\mathbf{u}$, and $\hat{\mathcal{H}}^l$ to
		obtain the fast-timescale parameters  $\{\mathbf{w}_k\}$,  $\{\mathbf{v}_b\}$ and $\{\mathbf{c}_k\}$.
	\end{algorithmic}
\end{algorithm}

In general, the average sum rate in (P3) serves as the fitness function for the PSO. However, it involves a large number of average sum rate calculations over a large set of random channel samples, which can make the computation prohibitively costly. To address this issue, we develop the RS-PSO algorithm, which efficiently evaluates the fitness function while maintaining satisfactory performance. Specifically, all \(\overline{L}\) random channel samples are divided into \(\overline{N}\) batches, with each batch containing \(L_{\mathrm{S}} = \overline{L}/\overline{N}\) samples (assumed to be an integer for convenience). At each iteration $i$, we introduce a \textit{recursive sampling surrogate function} \cite{32}, which is defined as 
\begin{align}
	&J^{(i)}(\mathbf{s}) = (1 -  \kappa ^{(i)}) J^{(i-1)} +\kappa ^{(i)}\nonumber\\
	& 
	\sum\limits_{k\in\mathcal{K}}\varrho_k  \sum_{l=(i-1)L_{\mathrm{S}}+1}^{iL_{\mathrm{S}}} \frac{R_k\left(\mathbf{s}, \mathbf{w}_k,\mathbf{v},\mathbf{c} ,\hat{\mathcal{H}}^l\right)}{L_{\mathrm{S}}},
	\label{20}
\end{align}
where \( \kappa ^{(i)} \) is an iteration-dependent constant that adjusts the weight of new samples in computing the average sum rate.  
To ensure compliance with constraints \eqref{M2}--\eqref{M3}, we introduce an adaptive penalty factor $\tau|\mathcal{Q}(\mathbf{s})|_{\mathrm{c}}$, which leads to
\begin{align}\label{ada}
	\bar{J}^{(i)}(\mathbf{s})=
		J^{(i)}(\mathbf{s}) +
	\tau|\mathcal{Q}(\mathbf{s})|_{\mathrm{c}},
\end{align}
where \(\tau\) is a positive penalty parameter, and \(\mathcal{Q}(\mathbf{s})\) represents the penalty term for infeasible particles. Specifically, each element in \(\mathcal{Q}(\mathbf{s})\) corresponds to position-rotation pair \((\mathbf{q},\mathbf{u})\) of PA subarrays that violates constraints \eqref{M2} \cite{10818453, ding2024movable} or \eqref{M3} \cite{6dmatwc}. Thus, \(\mathcal{Q}(\mathbf{s})\) is defined as  
\begin{align}\label{pr}
	&\mathcal{Q}(\mathbf{s})=\{(\mathbf{q}_n,{\mathbf{q}}_b)|
	\|{\mathbf{q}}_{n}-
	{\mathbf{q}}_{b}\|^2< d_{\min}, 1 \leq n<b \leq B\} \nonumber\\
	&\cup
	\{(\mathbf{q}_b,\mathbf{u}_n)|\mathbf{n}
	(\mathbf{u}_n)^T
	(\mathbf{q}_{b}-\mathbf{q}_n)>  0,~ 1\leq n,b\leq B, n \neq b \} \nonumber\\
	&\cup \{(\mathbf{q}_n,\mathbf{u}_n)|
	\mathbf{n}(\mathbf{u}_n)^T\mathbf{q}_n< 0,~1\leq n\leq B\}.
\end{align}

At iteration \(i\), the \(j\)-th particle updates its position \(\mathbf{s}_j^{(i)}\) based on its velocity \(\mathbf{m}_j^{(i)}\). Let \(\hat{\mathbf{s}}_j\) be the local best position for particle $j$ and \(\mathbf{s}_{\mathrm{g}}\) the global best position across all particles. The velocity and position updates are respectively given by
\begin{align}
	\mathbf{m}_j^{(i)}&=\omega\mathbf{m}_j^{(i-1)}+
	c_1\tau_1(\hat{\mathbf{s}}_{j}^{(i-1)}
	-\mathbf{s}_j^{(i-1)})\nonumber\\
	&+c_2\tau_2(\mathbf{s}_{\mathrm{g}}^{(i-1)}
	-\mathbf{s}_j^{(i-1)}),\label{gb}\\
		\mathbf{s}_j^{(i)}&=\mathcal{G}(\mathbf{s}_j^{(i-1)}+
	\mathbf{m}_j^{(i)}),\label{vb}
\end{align}
where \(c_1\) and \(c_2\) are learning factors, \(\tau_1\) and \(\tau_2\) are uniform random variables in \([0,1]\), and \(\omega\) is the inertia weight. The projection function \(\mathcal{G}(\cdot)\) ensures feasibility w.r.t. constraint \eqref{M1}:  
\begin{align}
	[\mathcal{G}({\mathbf{s}_j^{(i)}})]_j &= \begin{cases} 
		-\frac{A}{2}, & \text{if } [\mathbf{s}_j^{(i)}]_j < -\frac{A}{2}, \\ 
		\frac{A}{2}, & \text{if } [\mathbf{s}_j^{(i)}]_j > \frac{A}{2}, \\ 
		[\mathbf{s}_j^{(i)}]_j, & \text{otherwise},
	\end{cases}
	\label{gg}
\end{align}
for \(1 \leq j \leq 3B\), where \(A\) is the side length of the BS region $\mathcal{C}$, modeled as a cube for simplicity.

In summary, our proposed two-timescale polarforming, position, and rotation (TT-PPR) optimization algorithm is outlined in Algorithm 3, in which the complexity of the slow-timescale optimization in Step 1 for solving problem (P3) is given by \(\mathcal{O}(SL_{\mathrm{S}}NB)\).
\vspace{-3pt}
\section{Simulation Results}
\vspace{-3pt}
In the simulation, we set \( N = 4 \), \( B = 16 \), $K=30$, and $\mathcal{C}$ as a cube with $A=1$ meter (m) sides. 
Users are situated within a 3D coverage area, which forms a spherical annulus ranging from $20$ m to $200$ m in radial distance from the BS center. The carrier frequency is 24 GHz. The inter-antenna spacing on each PA subarray is $\frac{\lambda}{2}$ and the minimum inter-subarray distance is \(d_{\min}=\frac{\sqrt{2}}{2}\lambda + \frac{\lambda}{2}\). In Algorithm 3, we set $S = 200$, $\kappa ^{(i)} ={i^{-0.2}}$, and $I_{\mathrm{iter}} = 100$. We set the number of channel samples $\overline{L}=4000$ with $L_{\mathrm{S}}=40$, unless specified otherwise.  

To start with, we evaluate the performance of the proposed polarforming-based localization algorithm and compare it with the direct positioning method in \cite{24p}. The direct positioning method sets \(L = T_{\mathrm{s}}\) with \(P = 1\) and applies gradient‐based method to iteratively optimize the users' DoA vectors \(\mathbf{f}_k\), user-to-BS distances \(d_k\), and dynamic polarformed channels \(\eta_{m,k}\) to maximize the log-likelihood function. The localization error is calculated by $\sqrt{\mathbb{E}\left\{ \| \hat{\mathbf{p}} - \mathbf{p} \|^2 \right\}}
$ \cite{ref28},
where $\mathbf{p}$ represents the true locations of all users, and $\hat{\mathbf{p}}$ denotes their estimated locations. 
During the localization phase, we design \( \mathbf{X} \) as a semi-unitary matrix satisfying \( \mathbf{X}^H \mathbf{X} = \mathbf{I}_K \) and set \( \mathbf{w}_{k,p} \) based on the Fourier
transform matrix. We model the effective antenna gain in \eqref{gm} using  the 3GPP standard \cite{6dmatwc}.
\begin{figure*}[htbp]
	\vspace{-0.69cm}
	\begin{minipage}[t]{0.32\linewidth}
		\centering
		\includegraphics[width=2.60in]{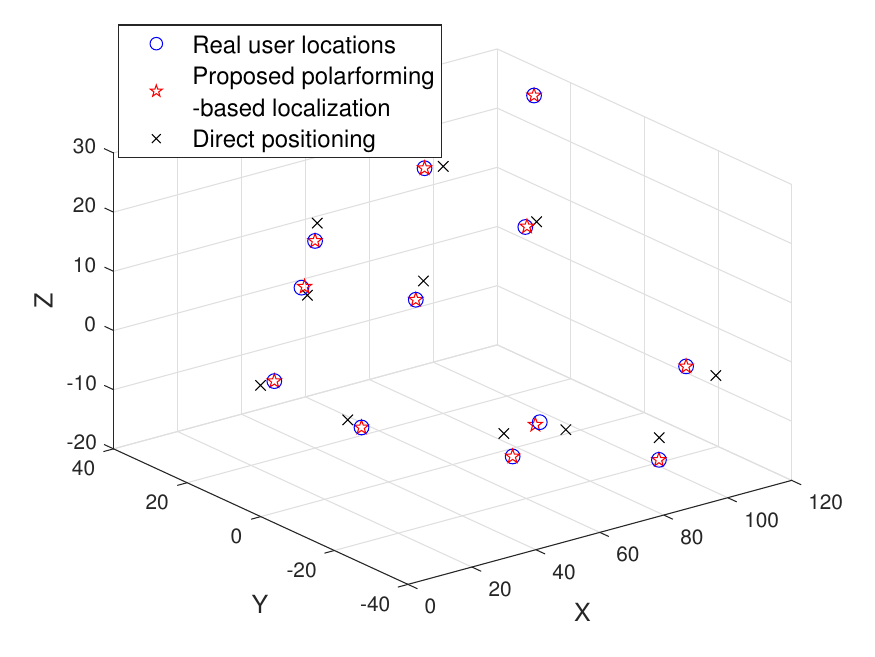}
		\setlength{\abovecaptionskip}{-8pt}
		\setlength{\belowcaptionskip}{-15pt}
		\caption{Results of user localization.}
		\label{local}
	\end{minipage}%
	\hspace{0.03in}
	\begin{minipage}[t]{0.33\linewidth}
		\centering
		\includegraphics[width=2.6in]{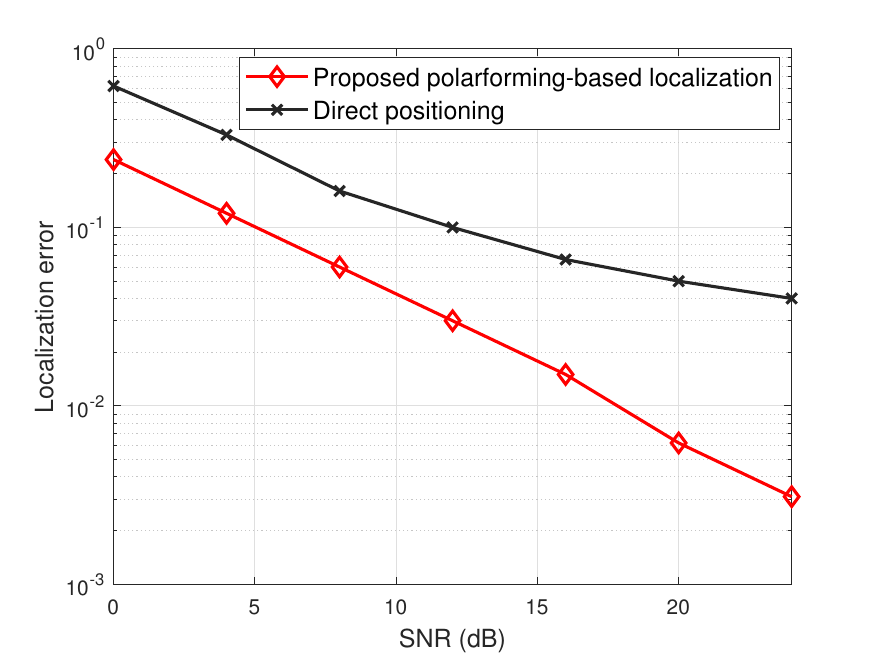}
		\setlength{\abovecaptionskip}{-8pt}
		\setlength{\belowcaptionskip}{-15pt}
		\caption{Localization error vs. the received SNRs in dB.}
		\label{error}
	\end{minipage}%
	\hspace{0.03in}
	\begin{minipage}[t]{0.32\linewidth}
		\centering
		\includegraphics[width=2.60in]{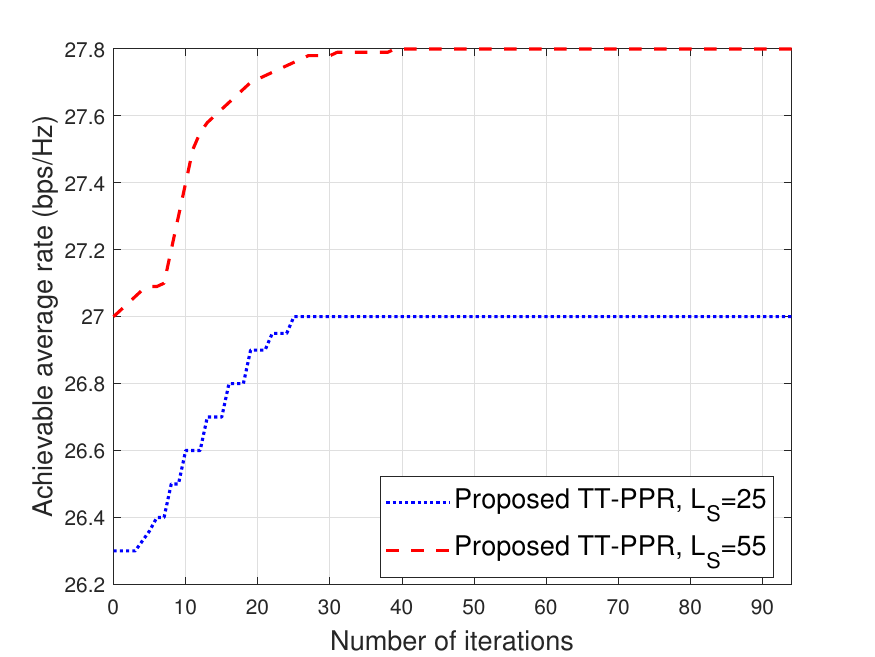}
		\setlength{\abovecaptionskip}{-8pt}
		\setlength{\belowcaptionskip}{-15pt}
		\caption{Impact of mini-batch size on the TT-PPR algorithm performance.}
		\label{paiter}
	\end{minipage}
	\vspace{-0.2cm}
\end{figure*}

First, the estimated locations for one localization realization of all users are shown in Fig. \ref{local}. Blue circles represent the actual user locations, while the red stars and black crosses denote the estimates from the proposed polarforming-based localization and direct positioning methods, respectively. The red stars closely align with the blue circles, whereas the black crosses show clear deviations, which indicates that the proposed method achieves higher accuracy than direct positioning. 
This is because, by providing controllable polarforming vectors, the proposed method fully exploits the measurement diversity of the received signal to improve estimation accuracy. Moreover, the proposed method efficiently exploits the Khatri-Rao channel structure to separate the estimation of stable channels \(\mathbf{H}_m\) from dynamic coefficients \(\boldsymbol{\Omega}_m\), thereby enabling user localization using the decoupled estimates with closed-form solutions, without the need to estimate \(\eta_{m,k}\) explicitly. In contrast, the direct positioning method neglects the Khatri-Rao structure of the composite channel and has to jointly estimate all user positions and channel coefficients in a larger coupled parameter space that lacks closed-form solutions, which makes it more prone to estimation errors and local optima.

Fig. \ref{error} illustrates the localization error (in terms of MSE between the actual and estimated user locations) of both the proposed and benchmark schemes versus (vs.) received signal-to-noise ratio (SNR). Fig. \ref{error} shows that increasing the SNR leads to a significant reduction in localization error for both schemes. Moreover, the localization error of the proposed scheme is less than 0.1 m for SNR larger than 5 dB, which is much lower than that of the direct positioning method. This
implies that the effective data rates of users can be significantly
increased as less sensing time is required for localization.

Next, we evaluate the performance of the proposed TT-PPR algorithm for joint polarforming and antenna position/rotation optimization for users' average sum-rate maximization by considering the following special cases of the proposed  algorithm:
\begin{itemize}
	\item \textbf{Fixed Parameter Scheme}:
In this scheme, \( \{\mathbf{w}_k\} \), \( \{\mathbf{v}_b\} \), \( \mathbf{q} \), and \( \mathbf{u} \) are fixed. We adopt a three-sector BS configuration, modeled as a special case of the PA system with \( B = 3 \) sectors. Each sector has approximately \(\lceil\frac{NB}{3}\rceil\) antennas covering \(120^\circ\). The precoding  vector \( \{\mathbf{c}_k\} \) uses maximum-ratio transmission (MRT), while the phase shifts and amplitudes of the polarforming vectors \( \{\mathbf{w}_k\} \) and \( \{\mathbf{v}_b\} \) are randomly generated and then fixed.

	\item \textbf{Proposed Polarforming Optimization Only}:
		Optimize only the polarforming vectors \( \{\mathbf{w}_k\} \) and \( \{\mathbf{v}_b\} \) using Algorithm 2, while keeping the positions, rotations, and precoding vectors  the same as in the fixed parameter scheme.
		
	\item \textbf{Proposed Position-Rotation Optimization Only}:
Optimize only the antenna position and rotation parameters \( \mathbf{q} \) and \( \mathbf{u}\) using Algorithm 3, while keeping all other parameters the same as in the fixed parameter scheme.
\end{itemize}

Fig. \ref{paiter} shows the impact of different mini-batch sizes on the convergence of the proposed TT-PPR algorithm.
It is observed that the TT-PPR algorithm can converge within 100 iterations under different $L_{\mathrm{S}}$ values. Although larger batch sizes for \( L_{\mathrm{S}} \) achieve a higher average rate over iterations compared to smaller batch sizes, they also incur higher computational complexity.

\begin{figure*}[htbp]
	\vspace{-0.69cm}
	\begin{minipage}[t]{0.32\linewidth} 
		\centering
		\includegraphics[width=2.60in]{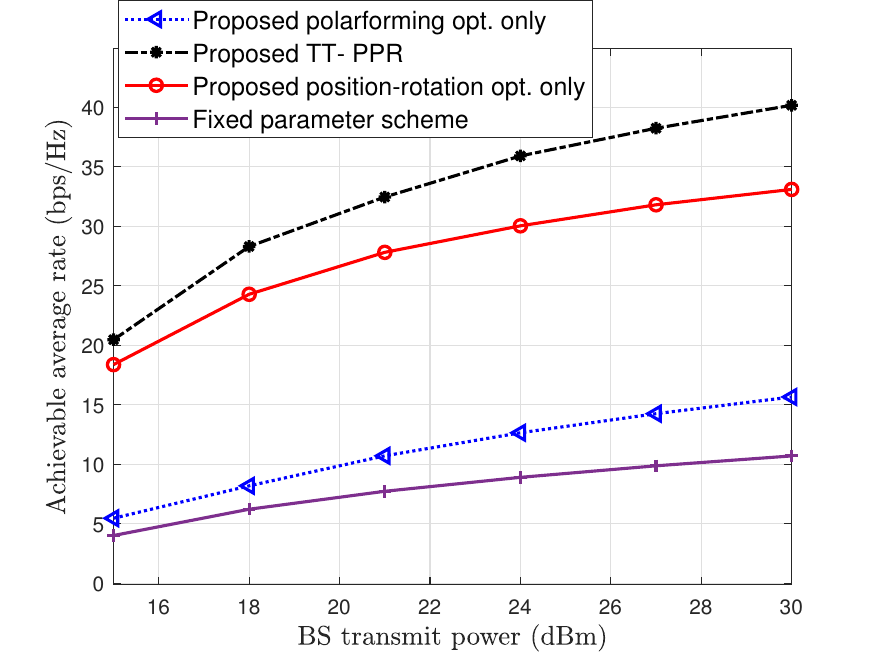}
		\setlength{\abovecaptionskip}{-8pt}
		\setlength{\belowcaptionskip}{-15pt}
		\caption{Achievable average rate  vs. BS transmit power for different schemes.}
		\label{papower}
	\end{minipage}
	\hspace{0.03in} 
	\begin{minipage}[t]{0.32\linewidth}
		\centering
		\includegraphics[width=2.578in]{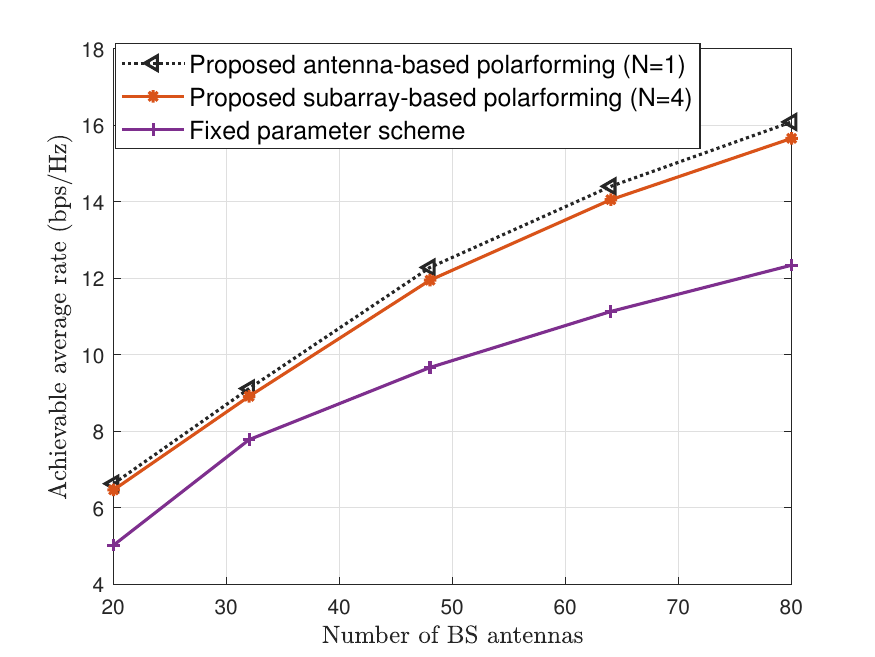}
		\setlength{\abovecaptionskip}{-8pt}
		\setlength{\belowcaptionskip}{-15pt}
		\caption{Achievable average rate vs. number of BS antennas for antenna- and subarray-based polarforming methods.}
		\label{subarray}
	\end{minipage}%
	\hspace{0.03in} %
	\begin{minipage}[t]{0.32\linewidth}
		\centering
		\includegraphics[width=2.60in]{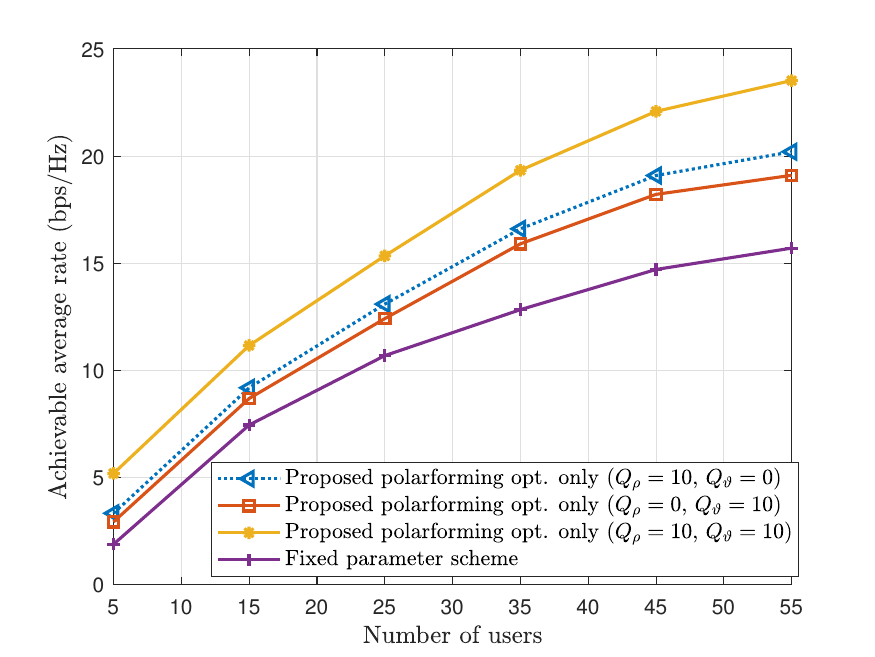}
		\setlength{\abovecaptionskip}{-8pt}
		\setlength{\belowcaptionskip}{-15pt}
		\caption{Achievable average rate vs. number of users for different polarforming quantization levels.}
		\label{bit}
	\end{minipage}
		\vspace{-0.18cm}
\end{figure*}

In Fig. \ref{papower}, we plot the achievable rates by different schemes versus BS's transmit
power. The results demonstrate that any level of antenna reconfiguration, i.e., proposed polarforming optimization only, position-rotation optimization only, and TT-PPT algorithm, can increase the achievable rate compared to the fixed parameter scheme, while joint antenna position/rotation and polarforming optimization by the TT-PPT
algorithm achieves the highest
performance gain. In particular, even the proposed polarforming optimization only scheme achieves a significant gain in achievable average rate
over the fixed parameter scheme. This is due to the ability of polarforming antenna
to dynamically adjust the polarization of the users and BS to maximize the instantaneous channel gain, thus effectively leveraging the additional DoFs provided by
polarization diversity. Moreover, it is observed
that with the same transmit power, the position-rotation optimization only scheme can also
achieve higher rates than the fixed parameter scheme. This is because the PA system with position-rotation adjustment  has
more spatial DoFs and can deploy the antenna resources more reasonably to match the users' channel spatial distribution. 
Furthermore, the performance gap between the proposed schemes and the fixed-parameter scheme widens with increasing BS transmit power. 
This is expected as the sum rate becomes more interference-limited as the transmit power increases, and adjusting user/BS polarforming and the antenna positions/rotations at BS can effectively improve the multiuser-MIMO channel condition for interference suppression by BS's transmit precoding.

The effect of the number of antennas at the BS on
the proposed polarforming optimization only scheme is also shown in Fig. \ref{subarray}. We examine two cases in
which each PA subarray at the BS employs the same polarforming vector, i.e., the proposed subarray-based polarforming scheme with $N=4$,
or each antenna at the BS can have independent polarization
control, i.e., the proposed antenna-based polarforming scheme with $N=1$.
We observe that even though the subarray-based polarforming has less flexibility in polarforming configuration, its performance is very close to that of the antenna-based polarforming, which, however,  requires higher hardware costs. This demonstrates that the proposed subarray-based polarforming scheme provides a more practically appealing balance between performance and hardware cost.  Furthermore, Fig. \ref{subarray} also shows that the performance advantage of polarforming becomes increasingly significant compared to the fixed parameter scheme as the number of BS antennas increases. This is because, with a larger number of BS antennas, the PA-enhanced system possesses more polarization DoFs, which allows polarization reconfiguration to better utilize these DoFs to optimize channel conditions and improve achievable rates.

 In Fig. \ref{bit}, we investigate the achievable average sum-rate of the proposed polarforming optimization-only scheme versus the number of users, \(K\). The effect of different polarforming amplitude/phase quantization levels on the achievable rate is also shown. The number of quantization bits is assumed to be the same at both the BS and users.
It can be observed that polarforming optimization with both amplitude and phase control outperforms both phase-only control (i.e., $Q_{\rho}=0$) and amplitude-only control (i.e., $Q_{\vartheta}=0$). Moreover, the proposed scheme with amplitude control achieves better performance than phase-only control. Furthermore, the performance gain achieved by amplitude control becomes more pronounced as the number of users, \(K\), increases, since multiuser interference due to CSI estimation errors becomes more severe. This result suggests that in systems with a large number of users, it may be more beneficial to employ amplitude-phase joint control polarforming rather than amplitude/phase-only polarforming.
\vspace{-3pt}
\section{Conclusion}
\vspace{-3pt}
In this paper, we have proposed a novel approach to enhance wireless sensing and communication performance by developing PA, which intelligently adjusts its polarization through polarforming vector and adaptively tunes its rotation and position to cater to channel variations.
Specifically, we have modeled the PA channel and designed an efficient polarforming-based user localization algorithm that extracts the common parameters in the stable unpolarformed channel across different controllable polarforming vectors.
Then, given the practical movement constraints of PAs, the transceiver polarforming and the antenna positions/rotations were jointly optimized
to maximize the weighted sum rate of users in an PA-enhanced ISAC
system. By applying the PDD and recursive sampling surrogate
techniques, an efficient algorithm has been proposed to balance the system performance and computational complexity.
Extensive simulation results under various practical setups demonstrated the effectiveness of localization, polarforming, and antenna placement. PA can significantly enrich the design DoFs for sensing and communication by tuning polarforming vectors and antenna positions/rotations at a limited cost, thereby opening up promising new applications and research directions for future wireless networks.
\vspace{-7pt}
\begin{appendices}
\section{Solution to (P2-3)}\label{App:SimuLoS}
The remaining variables in problem (P2-3) are optimized by solving the following subproblems.

1) {Optimization of $\mathbf{v}_b$:}
The update of the BS polarforming vectors $\{\mathbf{v}_b\}$ can be conducted by solving the following unconstrained QP problem:
\begin{align}	\label{hv1}
	\text{(P2-3.3) :}~~&~
	\mathop{\min}\limits_{\{\mathbf{v}_b\}}~~\sum\limits_{k\in\mathcal{K}}\varrho_k\epsilon_k|\xi_k|^2 \sum_{j\in\mathcal{K} }\bigg|\sum\limits_{b\in\mathcal{B}}{\varepsilon}_{k,j,b}\mathbf{v}_b^H\widehat{\mathbf{d}}_{k,b}\bigg|^2\nonumber\\
	&-\sum\limits_{k\in\mathcal{K}}\varrho_k\epsilon_k2 \operatorname{Re} \big\{ \xi_k^* \sum\limits_{b\in\mathcal{B}}\varepsilon_{k,k,b}\mathbf{v}_b^H\widehat{\mathbf{d}}_{k,b}\big\}\nonumber\\
	&
	+\frac{1}{2\mu}\sum\limits_{b\in\mathcal{B}} \left\| \mathbf{v}_b - \overline{\mathbf{v}}_b + \mu \bar{\mathbf{t}}_b \right\|^2,
\end{align}
where $\widehat{\mathbf{d}}_{k,b}=\mathbf{A}_{k,b}\mathbf{w}_k$ and $\varepsilon_{k,j,b}=(\mathbf{h}_{k,b}^{\mathrm{LoS}})^T\mathbf{c}_{j,b}$ with  $\mathbf{c}_{k,b}\in\mathbb{C}^{N\times 1}$ drawn from  \(\mathbf{c}_k = [\mathbf{c}_{k,1}^T,  \dots, \mathbf{c}_{k,B}^T]^T\).
The optimal solution of (P2-3.3) can be expressed as
\begin{align}
	\mathbf{v}_b^{\mathrm{opt}} = \bar{\mathbf{C}}_b^{-1} \bar{\mathbf{\mathbf{b}}}_b,
\end{align}
where
\begin{align}
	&\!\!\!	\!\!\bar{\mathbf{C}}_b\! =\! \sum_{k \in \mathcal{K}} \varrho_k \epsilon_k |\xi_k|^2\sum_{j \in \mathcal{K}}2 \varepsilon_{k,j,b} \varepsilon_{k,j,b}^* \widehat{\mathbf{d}}_{k,b} \widehat{\mathbf{d}}_{k,b}^H \!+\! \frac{1}{\mu} \mathbf{I}, \\
	&\!\!\!	\bar{\mathbf{b}}_b\! = \!\sum_{k \in \mathcal{K}} \varrho_k \epsilon_k 2 \xi_k^* \varepsilon_{k,k,b} \widehat{\mathbf{d}}_{k,b} + \frac{1}{\mu} \left( \overline{\mathbf{v}}_b - \mu \bar{\mathbf{t}}_b \right).
\end{align}

2) {Optimization of $\overline{\mathbf{v}}_b$:}
The $\overline{\mathbf{v}}_b$-subproblem is given by 
\begin{subequations}
	\label{ovv}
	\begin{align}
		\text{(P2-3.4) :}~\min_{\overline{\mathbf{v}}_b}~&~\|\mathbf{v}_b-\overline{\mathbf{v}}_b  + \mu \bar{\mathbf{t}}_b\|^2  \\
		\text {s.t.}~&~[\overline{\mathbf{v}}_b]_i \in \mathcal{F}, \quad \forall i \in \{1, 2\}.
	\end{align}
\end{subequations}
Similar to problem (P2-3.2), the optimal solution of problem (P2-3.4) can be efficiently computed in parallel, and its expression is omitted for brevity.

3) {Optimization of $\xi_k$:} With all other variables fixed, minimizing $\sum_{k \in \mathcal{K}} \varrho_k\epsilon_ke_k$ leads to the following linear minimum mean square error (LMMSE) equalizer coefficient
\begin{align}
	\xi_k=\frac{\mathbf{h}_k(\mathbf{q},\mathbf{u},\mathbf{w}_k,\mathbf{v})^H
		\mathbf{c}_k}{\sum_{j\in\mathcal{K} }|\mathbf{h}_k(\mathbf{q},\mathbf{u},\mathbf{w}_k,\mathbf{v})^H
		\mathbf{c}_j|^2+\sigma^2}.
\end{align}

4) {Optimization of $\epsilon_k$:}  The optimal solution is  $\epsilon_k = \frac{1}{e_k}$.

5) {Optimization of $\mathbf{c}$:} The transmit precoder update is obtained by solving:
\begin{subequations}
	\label{poww}
	\begin{align}
		\text{(P2-3.5) :}~
		\mathop{\min}\limits_{\mathbf{c}}~&~\sum\limits_{k\in\mathcal{K}}\varrho_k\epsilon_ke_k
		\\
		\text {s.t.}~&~\eqref{pow1}.
	\end{align}
\end{subequations}
Since \(\mathbf{c}_k\) terms are decoupled in the Lagrangian function, the optimal solution can be derived using the following first-order optimality condition: 
\begin{align}
	\mathbf{c}_k(\tilde{\mu}) =\varrho_k \epsilon_k \xi_k^* \Bigg( \tilde{\mu} \mathbf{I} + \sum_{j \in \mathcal{K}} \varrho_j \epsilon_j |\xi_j|^2 \mathbf{h}_j \mathbf{h}_j^H \Bigg)^{-1}  \mathbf{h}_k,
\end{align}
where \(\tilde{\mu}\) is the dual variable for the transmit power constraint. If \(\sum_{k \in \mathcal{K}} \|\mathbf{c}_k(0)\|^2 \leq \zeta\), then \(\{\mathbf{c}_k(0)\}\) is optimal; otherwise, the optimal \(\tilde{\mu}\) can be found via the bisection method \cite{ming0}. 
The derivation of the solution to problem (P2-3) is thus complete.
\end{appendices}
\vspace{-5pt}
\bibliographystyle{IEEEtran}
\bibliography{fabs}

\begin{thebibliography}{10}
\providecommand{\url}[1]{#1}
\csname url@samestyle\endcsname
\providecommand{\newblock}{\relax}
\providecommand{\bibinfo}[2]{#2}
\providecommand{\BIBentrySTDinterwordspacing}{\spaceskip=0pt\relax}
\providecommand{\BIBentryALTinterwordstretchfactor}{4}
\providecommand{\BIBentryALTinterwordspacing}{\spaceskip=\fontdimen2\font plus
\BIBentryALTinterwordstretchfactor\fontdimen3\font minus
  \fontdimen4\font\relax}
\providecommand{\BIBforeignlanguage}[2]{{%
\expandafter\ifx\csname l@#1\endcsname\relax
\typeout{** WARNING: IEEEtran.bst: No hyphenation pattern has been}%
\typeout{** loaded for the language `#1'. Using the pattern for}%
\typeout{** the default language instead.}%
\else
\language=\csname l@#1\endcsname
\fi
#2}}
\providecommand{\BIBdecl}{\relax}
\BIBdecl

\bibitem{isac1}
F.~Liu, Y.-F. Liu, A.~Li, C.~Masouros, and Y.~C. Eldar, ``Cramér-rao bound
  optimization for joint radar-communication beamforming,'' \emph{IEEE Trans.
  Signal Process.}, vol.~70, pp. 240--253, Jan. 2022.

\bibitem{isac2}
X.~Song, J.~Xu, F.~Liu, T.~X. Han, and Y.~C. Eldar, ``Intelligent reflecting
  surface enabled sensing: Cramér-rao bound optimization,'' \emph{IEEE Trans.
  Signal Process.}, vol.~71, pp. 2011--2026, Jun. 2023.

\bibitem{wangzhe}
Z.~Wang \emph{et~al.}, ``Extremely large-scale {MIMO}: Fundamentals,
  challenges, solutions, and future directions,'' \emph{IEEE Wireless Commun.},
  pp. 1--9, Mar. 2023.

\bibitem{haiquan1}
H.~Lu and Y.~Zeng, ``Communicating with extremely large-scale array/surface:
  Unified modeling and performance analysis,'' \emph{IEEE Trans. Wireless
  Commun.}, vol.~21, no.~6, pp. 4039--4053, Jun. 2022.

\bibitem{pol1}
Y.~He, X.~Cheng, and G.~L. Stuber, ``On polarization channel modeling,''
  \emph{IEEE Wireless Commun.}, vol.~23, no.~1, pp. 80--86, Feb. 2016.

\bibitem{duala}
T.~Kim, B.~Clerckx, D.~J. Love, and S.~J. Kim, ``Limited feedback beamforming
  systems for dual-polarized {MIMO} channels,'' \emph{IEEE Trans. Wireless
  Commun.}, vol.~9, no.~11, pp. 3425--3439, Nov. 2010.

\bibitem{qingisac}
M.~Hua, Q.~Wu, W.~Chen, and A.~L. Swindlehurst, ``Secure intelligent reflecting
  surface-aided integrated sensing and communication,'' \emph{IEEE Trans.
  Wireless Commun.}, vol.~23, no.~1, pp. 575--591, Jan. 2024.

\bibitem{proc}
Q.~Wu \emph{et~al.}, ``Intelligent surfaces empowered wireless network: Recent
  advances and the road to {6G},'' \emph{Proc. IEEE}, vol. 112, no.~7, pp.
  724--763, Jul. 2024.

\bibitem{6dmatwc}
X.~Shao, Q.~Jiang, and R.~Zhang, ``{6D} movable antenna based on user
  distribution: Modeling and optimization,'' \emph{IEEE Trans. Wireless
  Commun.}, vol.~24, no.~1, pp. 355--370, Jan. 2025.

\bibitem{6dmaDis}
X.~Shao, R.~Zhang, Q.~Jiang, and R.~Schober, ``{6D} movable antenna enhanced
  wireless network via discrete position and rotation optimization,''
  \emph{IEEE J. Sel. Areas Commun.}, vol.~43, no.~3, pp. 674--687, Mar. 2025.

\bibitem{6dmaMag}
X.~Shao and R.~Zhang, ``{6DMA} enhanced wireless network with flexible antenna
  position and rotation: Opportunities and challenges,'' \emph{IEEE Commun.
  Mag.}, vol.~63, no.~4, pp. 121--128, Apr. 2025.

\bibitem{poa}
S.-C. Kwon and A.~F. Molisch, ``Capacity maximization with polarization-agile
  antennas in the {MIMO} communication system,'' in \emph{IEEE Global Commun.
  Conf.}, Dec. 2015, pp. 1--6.

\bibitem{shao2025tutorial}
X.~Shao, W.~Mei, C.~You \emph{et~al.}, ``A tutorial on six-dimensional movable
  antenna for {6G} networks: Synergizing positionable and rotatable antennas,''
  \emph{arXiv preprint arXiv:2503.18240}, 2025.

\bibitem{jiang2025statistical}
Q.~Jiang \emph{et~al.}, ``Statistical channel based low-complexity rotation and
  position optimization for {6D} movable antennas enabled wireless
  communication,'' \emph{arXiv preprint arXiv:2504.20618}, 2025.

\bibitem{pi20246d}
X.~Pi, L.~Zhu, H.~Mao, Z.~Xiao, X.-G. Xia, and R.~Zhang, ``{6D} movable antenna
  enhanced multi-access point coordination via position and orientation
  optimization,'' \emph{arXiv preprint arXiv:2412.10736}, 2024.

\bibitem{liu2024uav}
C.~Liu, W.~Mei, P.~Wang, Y.~Meng, B.~Ning, and Z.~Chen, ``{UAV}-enabled passive
  {6D} movable antennas: Joint deployment and beamforming optimization,''
  \emph{arXiv preprint arXiv:2412.11150}, 2024.

\bibitem{10918750}
T.~Ren, X.~Zhang, L.~Zhu, W.~Ma, X.~Gao, and R.~Zhang, ``{6D} movable antenna
  enhanced interference mitigation for cellular-connected {UAV}
  communications,'' \emph{IEEE Wireless Communications Letters}, pp. 1--1,
  2025.

\bibitem{ppp}
Z.~Zhou, J.~Ding, C.~Wang, B.~Jiao, and R.~Zhang, ``Polarforming for wireless
  communications: Modeling and performance analysis,'' \emph{arXiv preprint
  arXiv:2409.07771}, 2024.

\bibitem{shaos}
X.~Shao, C.~You, W.~Ma, X.~Chen, and R.~Zhang, ``Target sensing with
  intelligent reflecting surface: Architecture and performance,'' \emph{IEEE J.
  Sel. Areas Commun.}, vol.~40, no.~7, pp. 2070--2084, Jul. 2022.

\bibitem{nsr}
X.~Shao and R.~Zhang, ``Enhancing wireless sensing via a target-mounted
  intelligent reflecting surface,'' \emph{Nat. Sci. Rev.}, vol.~10, no.~8, p.
  nwad150, Jul. 2023.

\bibitem{yumeng}
Y.~Zhang \emph{et~al.}, ``Full-space wireless sensing enabled by multi-sector
  intelligent surfaces,'' \emph{arXiv preprint arXiv:2406.15945}, 2024.

\bibitem{liuan}
A.~Liu \emph{et~al.}, ``A survey on fundamental limits of integrated sensing
  and communication,'' \emph{IEEE Commun. Surv. Tutorials}, vol.~24, no.~2, pp.
  994--1034, Feb. 2022.

\bibitem{6dmac}
X.~Shao \emph{et~al.}, ``Distributed channel estimation and optimization for
  {6D} movable antenna: Unveiling directional sparsity,'' \emph{IEEE J. Sel.
  Top. Signal Process.}, vol.~19, no.~2, pp. 349--365, Mar. 2025.

\bibitem{airs}
H.~Lu, Y.~Zeng, S.~Jin, and R.~Zhang, ``Aerial intelligent reflecting surface:
  Joint placement and passive beamforming design with {3D} beam flattening,''
  \emph{IEEE Trans. Wireless Commun.}, vol.~20, no.~7, pp. 4128--4143, Jul.
  2021.

\bibitem{heap}
M.~R. Castellanos and R.~W. Heath, ``Linear polarization optimization for
  wideband {MIMO} systems with reconfigurable arrays,'' \emph{IEEE Trans.
  Wireless Commun.}, vol.~23, no.~3, pp. 2282--2295, Mar. 2024.

\bibitem{43}
Y.~Rong, M.~R.~A. Khandaker, and Y.~Xiang, ``Channel estimation of dual-hop
  {MIMO} relay system via parallel factor analysis,'' \emph{IEEE Trans.
  Wireless Commun.}, vol.~11, no.~6, pp. 2224--2233, Jun. 2012.

\bibitem{xinran}
X.~Zhang \emph{et~al.}, ``Sparsity-structured tensor-aided channel estimation
  for {RIS}-assisted {MIMO} communications,'' \emph{IEEE Commun. Lett.},
  vol.~26, no.~10, pp. 2460--2464, Oct. 2022.

\bibitem{limei}
L.~Hu \emph{et~al.}, ``Triple {IRS}-aided communications: Row-column sparsity
  enhanced bayesian tensor learning for channel estimation,'' \emph{IEEE Trans.
  Commun., early access}, May. 2025.

\bibitem{9805460}
X.~Shao, L.~Cheng, X.~Chen, C.~Huang, and D.~W.~K. Ng, ``Reconfigurable
  intelligent surface-aided {6G} massive access: Coupled tensor modeling and
  sparse bayesian learning,'' \emph{IEEE Trans. Wireless Commun.}, vol.~21,
  no.~12, pp. 10\,145--10\,161, Dec. 2022.

\bibitem{MUSIC}
P.~Gupta and S.~Kar, ``Music and improved music algorithm to estimate direction
  of arrival,'' in \emph{Int. Conf. Commun. Signal Process.}, Apr. 2015, pp.
  0757--0761.

\bibitem{38}
Q.~Shi, M.~Razaviyayn, Z.-Q. Luo, and C.~He, ``An iteratively weighted {MMSE}
  approach to distributed sum-utility maximization for a {MIMO} interfering
  broadcast channel,'' \emph{IEEE Trans. Signal Process.}, vol.~59, no.~9, pp.
  4331--4340, Sep. 2011.

\bibitem{pdd}
Q.~Shi and M.~Hong, ``Penalty dual decomposition method for nonsmooth nonconvex
  optimization—part {I}: Algorithms and convergence analysis,'' \emph{IEEE
  Trans. Signal Process.}, vol.~68, pp. 4108--4122, Jun. 2020.

\bibitem{10818453}
J.~Ding, Z.~Zhou, and B.~Jiao, ``Movable antenna-aided secure full-duplex
  multi-user communications,'' \emph{IEEE Trans. Wireless Commun.}, vol.~24,
  no.~3, pp. 2389--2403, Mar. 2025.

\bibitem{ding2024movable}
J.~Ding, Z.~Zhou \emph{et~al.}, ``Movable antenna-aided near-field integrated
  sensing and communication,'' \emph{arXiv preprint arXiv:2412.19470}, 2024.

\bibitem{con}
M.~Clerc and J.~Kennedy, ``The particle swarm - explosion, stability, and
  convergence in a multidimensional complex space,'' \emph{IEEE Trans. Evol.
  Comput.}, vol.~6, no.~1, pp. 58--73, Feb. 2002.

\bibitem{32}
A.~Liu, V.~K.~N. Lau, and M.-J. Zhao, ``Online successive convex approximation
  for two-stage stochastic nonconvex optimization,'' \emph{IEEE Trans. Signal
  Process.}, vol.~66, no.~22, pp. 5941--5955, Nov. 2018.

\bibitem{24p}
D.~Dardari, N.~Decarli, A.~Guerra, and F.~Guidi, ``{LOS/NLOS} near-field
  localization with a large reconfigurable intelligent surface,'' \emph{IEEE
  Trans. Wireless Commun.}, vol.~21, no.~6, pp. 4282--4294, Jun. 2022.

\bibitem{ref28}
M.~Ghermezcheshmeh and N.~Zlatanov, ``User localization via multiple
  intelligent reflecting surfaces for {LOS}-dominated channels,'' \emph{IEEE
  Access}, vol.~11, pp. 122\,446--122\,457, Oct. 2023.

\bibitem{ming0}
M.-M. Zhao \emph{et~al.}, ``Intelligent reflecting surface enhanced wireless
  networks: Two-timescale beamforming optimization,'' \emph{IEEE Trans.
  Wireless Commun.}, vol.~20, no.~1, pp. 2--17, Jan. 2021.

\end{thebibliography}
\vspace{-5pt}
\end{document}